\documentclass[aps,amsmath,amssymb,twocolumn,pra,10pt]{revtex4-2}

\usepackage{color}
\usepackage{physics}
\usepackage{amsthm}
\usepackage{bbm}
\usepackage{tikz}
\usepackage{xcolor}
\usepackage{soul}
\usepackage{hyperref}

\usepackage[compatibility=false]{caption}
\usepackage{graphicx}
\usepackage{dcolumn}
\usepackage{bm}

\usepackage{url}
\usepackage{subcaption}
\usepackage[justification=centering]{caption}

\newcommand{\la}[1]{{\cal L}(#1)}
\newcommand{\id}{\mathbb{I}}

\begin{document}

\title{Physically motivated decompositions of single-qutrit gates}

\author{Aryan Iliat$^1$}

\author{Mark Byrd$^1$}

\author{Sahel Ashhab$^{2,3}$}

\author{Lian-Ao Wu$^{4,5,6}$}

\affiliation{$^1$School of Physics and Applied Physics, Southern Illinois University, Carbondale, IL, 62903, USA}
\affiliation{$^2$Advanced ICT Research Institute, National Institute of Information and Communications Technology, 4-2-1, Nukui-Kitamachi, Koganei, Tokyo 184-8795, Japan}
\affiliation{$^3$Research Institute for Science and Technology, Tokyo University of Science, 1-3 Kagurazaka, Shinjuku-ku, Tokyo 162-8601, Japan}
\affiliation{$^4$ 
Department of Physics, 
University of the Basque Country UPV/EHU, 48080 Bilbao, Spain}
\affiliation{$^5$IKERBASQUE Basque Foundation for Science, 48013 Bilbao, Spain}
\affiliation{$^6$EHU Quantum Center,
University of the Basque Country UPV/EHU, Leioa, 48940, Spain}

\date{\today}

\begin{abstract}
Although only two quantum states of a physical system are often used to encode quantum information into qubits, in principle, qudits (a system with a number $d$ states) could be used to increase the information capacity of the system. In this work, $3$-state systems (qutrits) are studied using a parametrization of the group $U(3)$ that is relevant for controlling superconducting qutrits using fixed-frequency resonant control pulses.  The parametrization uses a decomposition of a general $3\times 3$ unitary matrix into the exponential of a diagonal matrix and the exponential of an off-diagonal matrix.  This decomposition is numerically confirmed to allow the parametrization of any element in $U(3)$.  However, many different sets of parameters can give the same element of $U(3)$.  Furthermore, different elements could have different numbers of decompositions.  Why are there different numbers?  Which decompositions are better and why?  These two questions are answered so that the parametrization can be better utilized and searches can be made more efficient.  Other parametrizations are derived and discussed and the methods used are applicable to these other parametrizations as well.  
\end{abstract}

\maketitle

\section{\label{sec:background}Introduction}

Quantum information processing with qudits (d-level quantum systems) has been considered since the early days of quantum information research \cite{PhysRevLett.70.1895, 10.3389/fphy.2020.589504, 10.1007/3-540-49208-9_27, Li_2013, Castelvecchi_2025}.  Indeed, in some cases, qudits can provide various advantages over qubits \cite{2014SCPMA..57.1712L, PhysRevLett.120.160502, PhysRevA.75.022313, PhysRevLett.94.230502, article}.  
More recently, the implementation of qudit gates in superconducting systems has received increasing interest, due to the rapid experimental advances that made such implementations possible \cite{doi:10.1142/S0129054103002011, PhysRevLett.105.223601, doi:10.1126/science.1173440, PhysRevLett.120.130503, Yurtalan_2021, PhysRevLett.110.120501, doi:10.1126/science.1134008, PhysRevX.11.021010, Ringbauer_2022}.

The optimal method to implement operations on a quantum computing device depends on the specifics of the physical system used to encode the quantum states \cite{Ladd_2010, Buluta_2011}. In particular, superconducting qubits are typically manipulated using fixed-frequency microwave control pulses \cite{PhysRevLett.130.260601, Bao:2024wwh}. An arbitrary unitary operation on a qubit is implemented by driving the qubit using a resonant pulse with a properly chosen amplitude and duration. The resonant control pulse results in a unitary operator that can be expressed as the exponential of an off-diagonal matrix. This operation can also be understood as being analogous to a rotation about an axis in the xy plane of the Bloch sphere representation of the qubit. In fact, they are between two different energy levels, so when restrict to those two, the picture is just the Bloch sphere picture.  The appropriate xy rotation, followed or preceded by the appropriate z-axis rotation, constitutes a decomposition that can produce any desired rotation of the Bloch sphere. This decomposition can be seen as a quantum version of Euler's decomposition of arbitrary rotations \cite{Nielsen}. In practice, the z-axis rotations are not implemented using separate control pulses. Instead, they are produced by appropriately shifting the phases of the resonant control pulses that implement the xy rotations. Inspired by this decomposition of qubit gates, Yurtalan {\it et al.} found and used a similar decomposition to implement the qutrit Walsh-Hadamard (WH) gate using a single control pulse with three frequency components that drive the three transitions in the qutrit \cite{PhysRevLett.125.180504}.  

Using this decomposition with arbitrary parameters, one can choose a randomly generated matrix from $U(3)$ and find a set of a parameters that give the matrix as discussed in Section \ref{sec:WH}.  In Ref.~\cite{PhysRevLett.125.180504}, this was done using a numerical search.  One observation that the authors of Ref.~\cite{PhysRevLett.125.180504} noted but could not explain was that the numerical search produces multiple solutions for the same unitary matrix. Furthermore, the number of different solutions depends on the given matrix and the different solutions may have no obvious relation between them.  This work is motivated by these findings.  The questions that are to be answered are the following:  Why are there different sets of parameters that give the same matrix?  How are these sets related?  Why do different matrices have different numbers of sets that give the same matrix? Are all the sets completely equivalent to each other, or do practical considerations favor some sets over others? The final question has direct implications for experiments. 

In this article, the decomposition of unitary matrices is referred to as the diagonal-off-diagonal decomposition is discussed in \ref{sec:WH} and, for completeness, some results found by Yurtalan {\it et al.} are reviewed.  This decomposition was shown to be capable of decomposing random unitaries.  But the number of possible sets of parameters depends on the matrix.  How this happens and how different sets of parameters that give the same matrix are related, are discussed \ref{sec:redsym}.  Since the WH matrix implements a Fourier transform operation and is therefore important for various protocols, a thorough discussion of this matrix is provided. By comparison with the shortest path to obtain a unitary matrix, the optimal paths of the diagonal-off-diagonal parametrization is found in section \ref{sec:opt}.  These results are important for experimental implementations of qutrit gates.  The number of different sets of parameters that give the same matrix, as well as the relationships between different sets, are shown to be related by symmetry arguments.

\section{\label{sec:WH}Parametrizations of SU(3)}

In this section, we discuss the decomposition, diagonal-off-diagonal, of Ref.~\cite{PhysRevLett.125.180504}.  This decomposition is motivated by experiments on superconducting qutrits, as discussed in the introduction.

In the diagonal-off-diagonal decomposition \cite{PhysRevLett.125.180504}, an element of $U(3)$ is decomposed as the exponential of a diagonal Hamiltonian and the exponential of an off-diagonal Hamiltonian.  Each of these can be expanded in terms of a complete set of $3\times 3$ Hermitian matrices, such as the Gell-Mann matrices, which is a commonly used set.  (See Appendix \ref{sec:AppendixA}.)  The diagonal unitary, denoted $U_d$, is the exponential of a linear combination of the identity matrix and the diagonal Gell-Mann matrices $\lambda_3$, and $\lambda_8$ \cite{Ramond_2010, Pfeifer2003}.  The off-diagonal matrix $U_o$ would be the exponential of a linear combination of the rest, so that 
\begin{equation}\label{eq:dod}
U = U_d U_o,
\end{equation}
where $U$ is any element of $U(3)$. 
Using the same notation as in \cite{PhysRevLett.125.180504}, Eq.~(\ref{eq:dod}) can be rewritten as 
$$U = e^{-iG_d}e^{-iG_o}$$
where 
$$G_d = \left(\begin{array}{ccc}
  \phi_0 & 0 & 0 \\
  0 & \phi_1 & 0 \\
  0 & 0 & \phi_2
\end{array}\right), \quad G_o = \left(\begin{array}{ccc}
  0 & m_{01} & m_{02} \\
  m_{01}^\ast & 0 & m_{12} \\
  m_{02}^\ast & m_{12}^\ast & 0
\end{array}\right),$$
are the Lie algebraic elements of the diagonal and off-diagonal $U(3)$ group, respectively. Note that the exponential of $G_d$ will contain only diagonal terms. However, the diagonal elements of the exponential of $G_o$ will not necessarily be zero even though the diagonal elements of $G_o$ are zero.  

This parametrization is experimentally motivated and has been numerically verified by choosing many random unitary matrices $U$ and finding parameter values $\{\phi_i, m_{jk}\}$ that will produce each matrix $U$. However, while this calculation shows sufficiency of this parametrization, it indicates that this is an ``over-parametrization" of the group because several distinct sets of parameters are found for each $U$. For example, for the Walsh-Hadamard gate, these parameter sets are shown in table \ref{tbl:tblS1}, which was originally presented in \cite{PhysRevLett.125.180504}.  Notice that there are five different sets. This multiplicity of decompositions is due to over-parametrization (a multiple cover of the group manifold).

\begin{table*}[ht]
	\caption{Numerically determined matrix elements of Walsh-Hadamard gate generators in the diagonal-off-diagonal decomposition explained in the text.  (Originally presented in \cite{PhysRevLett.125.180504}.)}
	\label{tbl:tblS1}
	\centering
	\begin{tabular*}{7in}{@{\extracolsep{\fill} } c c c c c c c }
		\hline
		\hline
		Decomposition & $m_{01}$ & $m_{12}$ & $m_{02}$ &$\phi_0$&$\phi_1$&$\phi_2$ \\
		\hline
		1&$-0.9672-0.2365i$&$1.9345$ &$-0.9672-0.2365i$& $0.8434$ & $0.3637$ & $0.3637$ \\
        \hline
        2&$-0.6982-1.2092i$&$1.3962$ &$-0.6981-1.2092i$& $1.9199$ & $6.1087$ & $6.1086$ \\
        \hline
		3&$-0.9672-1.6753i$&$0.6885+0.7194i$ &$0.2788-0.9559i$& $2.4581$ & $0.3637$ & $5.0322$ \\
        \hline
		4&$ 0.2788-0.9559i$&$0.6885-0.7194i$ &$-0.9672-1.6753i$& $2.4581$ & $5.0322$ & $0.3637$ \\
        \hline
    5&$0.3491+0.6046i$&$-0.6981$ &$0.3491+0.6046i$& $6.1086$ & $4.0143$ & $4.0143$ \\
		\hline 
		\hline
	\end{tabular*}
\end{table*}


 \section{Different sets of parameters: symmetry considerations}\label{sec:symm}

Suppose that a group element $U$ of $SU(N)$ is decomposed in to a product of $m$ operations, $U_i$.  Let this product be written as 
$$
U = U_1 U_2 U_3 ... U_m.
$$ 
Now also suppose that $U$ has a symmetry.  That is, suppose there exists an operation $R$ that leaves $U$ unchanged under conjugation, 
\begin{equation}\label{eq:rsym}
R U R^{-1} = U.  
\end{equation}
This implies that 
$$
U = R U_1 R^{-1} R U_2 R^{-1} R U_3 R^{-1} ... R U_m R^{-1}.
$$
However, it does not imply that $RU_i R^{-1} = U_i$ for any $U_i$.  Such relations may or may not hold.  This fact has important implications for parametrizations, and thus optimal evolutions, of quantum systems.  

Now consider the implications of this for the Walsh-Hadamard (WH) matrix and the parametrization introduced in Reference \cite{PhysRevLett.125.180504}.  For the parametrization $U_oU_d$, some $U$ have a different number of sets of parameters that will give the same specific gate. That is, there is a different number of sets of parameters that give the same matrix $U$.  Notice that in the Table \ref{tbl:tblS1}, the WH has five different decompositions.  As will be discussed, the number of different parametrizations depends on the symmetry of the matrix.  Some matrices have a much greater symmetry than others.

The WH matrix, which will be discussed further below, has the form 
$$
\frac{1}{\sqrt{3}}\left(\begin{array}{ccc}
  1 & 1 & 1 \\
  1 & e^{i\frac{2\pi}{3}} & e^{-i\frac{2\pi}{3}} \\
  1 & e^{-i\frac{2\pi}{3}} & e^{i\frac{2\pi}{3}}
\end{array}\right).
$$
Notice that this matrix is both unitary {\sl and} symmetric.  This extra symmetry implies that it has a greater degree of symmetry than a generic unitary matrix.  Therefore, there will be a larger set of operations that leave it invariant.  This leads to a greater number of possible ways to parameterize the matrix $V$.  This will be shown for the WH matrix in the next section. 

Note that different ways to express the same matrix leads to different possible ways to implement a physical gating operation.


\subsection{The dependence on different sets of parameters for different matrices}

Note that matrices with a higher degree of symmetry are rare in the space of all matrices $U$.  In fact, matrices with degeneracy have measure zero in the space, so that the probability of picking one of these matrices at random is zero.  However, if one chooses two Hamiltonians that generate two different $U$ and connects them using a convex combination, e.g., $H = xH_1 +(1-x)H_2$, with $0\leq x\leq 1$, the number of parametrizations can change abruptly as $x$ varies. Such an interpolation is used in adiabatic evolutions.  This is due to the fact that picking two points from the entire set and then connecting them by interpolating between them increases the probability of finding one of these rare points from zero to a significant amount.  This can be explained by the following simple, 2D example.  

Consider a square in the xy plane and draw a vertical line from top to bottom that divides the area in the square into two parts.  If one chooses a point at random in the area enclosed by the square.  The probability that this point lies along the line is zero.  However, if you choose two points inside the volume and connect them by a line, the chances that one of the points on the line lies on the line is significant.  For example, if the vertical line is in the center, the probability that a point along the connecting line contains a point along the vertical line is 1/2, which is equal to the probability that the two randomly chosen points lie in different halves of the square.  Here, the two points represent different Hamiltonians (corresponding to different unitaries) and the line between them is described by the parameter $x$ in the previous paragraph.  

\begin{figure*}[ht]
\centering
    
     \begin{subfigure}[t]{0.49\textwidth}
\includegraphics[width=\textwidth]{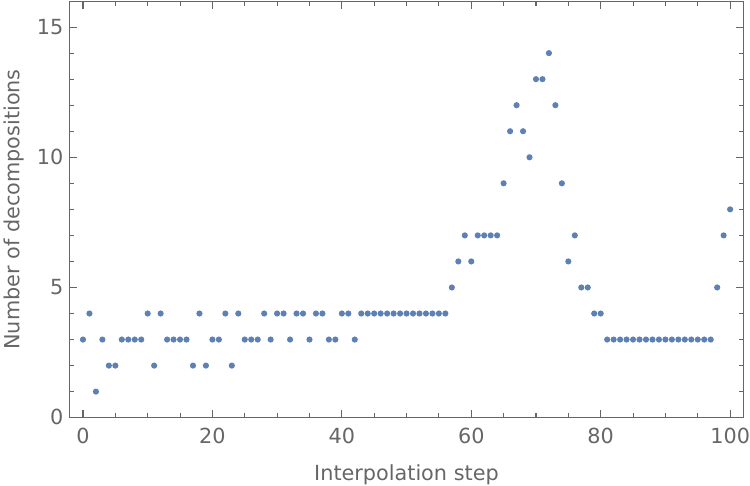}
         \caption{}
     \end{subfigure}
     \hfill
     \centering
     \centering
     \begin{subfigure}[t]{0.49\textwidth}
    \includegraphics[width=\textwidth]{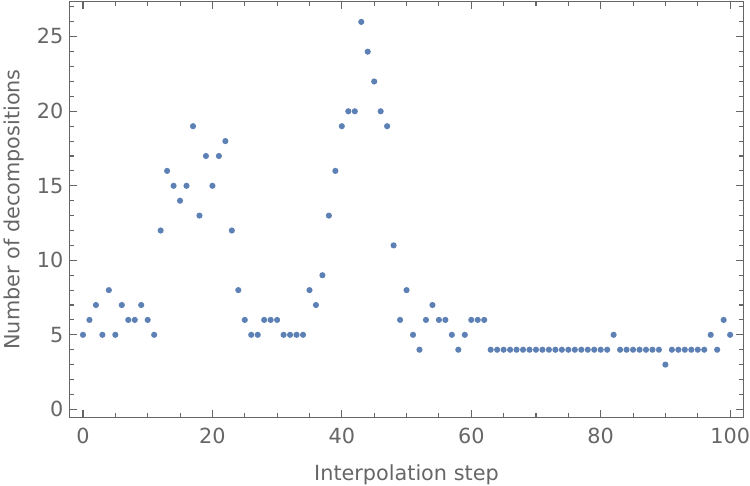}
         \caption{}
     \end{subfigure}
     \centering
    \caption{Number of decompositions as a function of step index as we gradually deform a unitary operator in $100$ steps, starting from one randomly generated unitary and ending with another randomly generated unitary. The two graphs correspond to two instances of the same calculation. The first and last data points in each graph represent two random unitaries $U_1$ and $U_2$. The generators ($G_1$ and $G_2$) of these two random unitaries are calculated. Then, using linear interpolation, we obtain 99 generators that gradually evolve from $G_1$ to $G_2$. Each intermediate generator is then used to generate a new unitary. As a result, each intermediate point in the graph represents a different unitary, and these unitaries gradually evolve from the first random unitary ($U_1$) to the other random unitary ($U_2$). The number of different decompositions, i.e.~the y-axis value, is obtained numerically by scanning the full range of decomposition parameters and identifying valid decompositions, as explained in the main text. For reference, the number of decompositions for the Walsh-Hadamard matrix is $5$.  Two random unitaries were used in (a), and two other unitaries were used in (b).}
        \label{fig:simu}
\end{figure*}

In other words, in the case of the Hamiltonian matrices, when we choose two random matrices and look for their parametrizations, the probability of choosing these special matrices, those with a high degree of symmetry, is zero.  However, when the two random matrices are connected by interpolation (as $x$ varies), the probability of crossing a high-symmetry region can be significant as shown by the example.  Explicitly calculating this probability can be difficult because of the high dimensionality of the space, but it is analogous to the two-dimensional example and is shown in Fig.~(\ref{fig:simu}).  

\subsection{Other possible parametrizations}

The diagonal-off-diagonal decomposition was chosen for experimental purposes.  However, there are other possible decompositions that can be derived by other methods.  One such method that has been utilized in the past is the Cartan decomposition (see for example 
\cite{helgason1979differential,hermann1966lie}).  This has been used for other parametrizations of $SU(3)$ \cite{ByrdSU3}, $SU(4)$ \cite{TilmaSU4}, and can be extended to $SU(N)$ as well as for describing time-optimal control of two-qubit gates \cite{KhanejaSU4}.  In Appendix \ref{sec:car}, several other parametrizations of $SU(3)$ are given that use the Cartan decomposition.  These are also experimentally motivated.  They allow for a parametrization that avoids two-photon transitions, which can be experimentally difficult to perform.  

Note that, since the parametrizations based on the Cartan decomposition are analytically obtained, they can also be parameterized such that the parameter ranges are (almost) uniquely identified.  However, this can be quite involved \cite{ByrdSU3}.  


\section{Relation between different sets of parameters for WH: Example 1}

In this section, the parameter sets presented in  Ref.~\cite{PhysRevLett.125.180504} for the WH matrix are shown to be related and that relationship is described by using the symmetry arguments above.  

Let $W$ be the Walsh-Hadamard matrix,
$$
W = \frac{1}{\sqrt{3}}\left(\begin{array}{ccc}
  1 & 1 & 1 \\
  1 & e^{i\frac{2\pi}{3}} & e^{-i\frac{2\pi}{3}} \\
  1 & e^{-i\frac{2\pi}{3}} & e^{i\frac{2\pi}{3}}
\end{array}\right),
$$
and let $S$ be the operator that leaves $\ket{0}$ unchanged and swaps states $\ket{1}$ and $\ket{2}$,
\begin{eqnarray}\label{eq1:smat}
S = \left(\begin{array}{ccc}
  1 & 0 & 0 \\
  0 & 0 & 1 \\
  0 & 1 & 0
\end{array}\right).
\end{eqnarray}
($S$ is an example of $R$ from Eq.~(\eqref{eq:rsym} above.)  Then we have the relations
\begin{eqnarray*}
\begin{cases}
SWS = W, \\
W^T = W, \\
W^2=S.
\end{cases}
\end{eqnarray*}

To see how the sets of parameters are related, note that if $U=\exp[iM]$ is a unitary matrix, then $\det U = \exp[i\mbox{Tr}M]$. Also, note that the determinant obeys the rule $\det(U_dU_o) = \det(U_d)\det(U_o)$.  In addition, the algebraic elements of $U_o$ are traceless, so $\det U = \det U_d =\exp[-i(\phi_0+\phi_1+\phi_2)]$. For the Walsh-Hadamard matrix, the determinant is $-i$, which means 
$\phi_0+\phi_1+\phi_2=\frac{\pi}{2}$.   Now, note that if each of these phases changes by $\frac{2\pi}{3}$ (such that the changes add up to $2\pi$), then the determinant of the unitary is unchanged.  A similar argument holds for $4\pi/3$.  Note that the center of $SU(3)$, which is the set of elements that commute with all elements of $SU(3)$, is $\{\id, \id e^{i2\pi/3},\id e^{i4\pi/3}\}$ \cite{Mallesh:1997cc, PhysRevD.111.034516}. 

Consider the symmetry matrix $S$ first.  The matrix $S$ will transform the set $3$ to the set $4$ from the Table \ref{tbl:tblS1}.  
In this transformation, the matrix $S$ will swap the elements $m_{01}$ and $m_{02}$ while  also exchanging $m_{12}$ and $m_{12}^*$.  It will also transform the phases $\phi_1$ and $\phi_2$ into each other, as shown by
\begin{widetext}
\begin{eqnarray}\label{eq:sw}
S\cdot\left(\begin{array}{ccc} 
0 & m_{01} & m_{02} \\ 
m_{01}^\ast & 0  & m_{12} \\ 
m_{02}^\ast & m_{12}^\ast & 0  \end{array}\right)\cdot S=
\left(\begin{array}{ccc}  0 & m_{02} & m_{01} \\ m_{02}^\ast & 0  & m_{12}^\ast \\ 
m_{01}^\ast & m_{12} & 0  \end{array}\right)
,\text{and}\quad
S\cdot\left(\begin{array}{ccc} 
\phi_0 & 0 & 0 \\ 
0 & \phi_1  & 0\\ 
0 & 0 & \phi_2   \end{array}\right)\cdot S=
\left(\begin{array}{ccc}  
\phi_0 & 0 & 0 \\ 
0 & \phi_2  & 0\\ 
0 & 0 & \phi_1  \end{array}\right),
\end{eqnarray}
\end{widetext}
which means that starting from set 3, we can get set 4 by applying the matrix transformation $S$.  
Note that $S$ is its own inverse, so that $SWS^\dagger =SWS =W$.  That is, $W$ is invariant under conjugation by $S$.  However, neither $U_o$ nor $U_d$ is invariant under conjugation by $S$.  This provides a way to see how a change in each of the two parts can give rise to the same matrix, $W$.


\subsection{Relation between different sets of parameters for WH: Example 2}

The other cases of Table \ref{tbl:tblS1} can be treated in a similar way, but will require different symmetry matrices, denoted $T$, such that $T W T^\dagger =W$.  This will provide different sets of parameters that give the WH matrix. The matrix $T$  
will be constructed such that it will represent all matrices that commute with $U$.  
As described in Sec.\ref{sec:symm}, the matrices $T$ will not necessarily commute with $U_o$ and $U_d$ individually.  When $T$ commutes with $U$, but not with the individual factors $U_o$ and $U_d$, the parameter changes in the matrix $U_d$ can compensate for the parameter changes in $U_o$, such that the matrix $U$ is unchanged. In this case, $U_d$ and $U_o$ are both changed, but $U$ is not.  In general, $T$ depends on $U$.  Matrices with a greater symmetry, will have a larger number of possible $T$ that satisfy these criteria.  If a greater symmetry is present, then the matrix is rare.  These matrices are rare due to the symmetry and there are a greater number of sets of parameters that give the matrix $U$ in the diagonal-off-diagonal decomposition.  

The strategy is to find all matrices, $T$, that commute with $U$, i.e., $T U T^\dagger = U$.  It is well-known that two matrices commute when they are diagonal in the same basis.  To find matrices that commute with a given matrix $U$, $U$ will be diagonalized.  In this basis, it will commute with any diagonal matrix. This strategy can be used to find out how all sets of parameters in Table \ref{tbl:tblS1} are related, just as was done in the last section for the special case where $T = S$.  Note that a matrix $U = e^{-iH}$ can be diagonalized by diagonalizing the Hermitian matrix $H$ since $M U M^{-1} = e^{-iM H M^{-1}}$.    

For the WH matrix, this method is explained in some detail here.  First, diagonalize the matrix $U$.  Suppose the matrix that diagonalizes it is $V$.  Then $VUV^\dagger = U_D$, is diagonal.  Then take the set of diagonal $T$, $T_D$ and transform to the $U$ basis, by conjugating the diagonal matrix by $V^{\dagger}$.  This will give $T= V^\dagger T_D V$, which commutes with $U$.


\subsection{\label{sec:redsym}Symmetries and Redundancies for the WH matrix}

 In this subsection, all matrices $T$ that commute with the WH matrix will be found.  This will allow the other sets of parameters in Table \ref{tbl:tblS1} to be related to each other.
 Recall that the method is to find the matrices $T$ that leave the WH matrix invariant under conjugation $T^\dagger W T = W$, but do not leave the separate parts $U_o$ and $U_d$ invariant.  This implies that $T$ will commute with the Walsh-Hadamard matrix $[W, T]=0$. Since $W$ and the matrix $T$ commute, they are simultaneously diagonalizable by the same matrix.  Assume $D$ is the diagonalizing matrix.  $D$ will diagonalize Walsh-Hadamard and its Hamiltonian. This means $DWD^\dagger=W_D$ and $D^\dagger T D = T_D$.  The diagonal form of $T$, $T_D$, will be written as $$T_D=\left(\begin{array}{ccc}
  e^{-i\theta_1} & 0 & 0 \\
  0 & e^{-i\theta_2} & 0 \\
  0 & 0 & e^{-i\theta_3}
\end{array}\right).$$ 
The parameters $\theta_1$, $\theta_2$ and $\theta_3$ determine the eigenvalues of $T$.  
The matrix $D$ is found by determining the eigenvalues of the Hamiltonian that generates the WH matrix.  It has the form
\begin{equation}
D=\left(
\begin{array}{ccc}
0 & -\frac{1}{\sqrt{2}} & \frac{1}{\sqrt{2}} \\ 
\frac{1 - \sqrt{3}}{\sqrt{(1 - \sqrt{3})^2 + 2}} & \frac{1}{\sqrt{(1 - \sqrt{3})^2 + 2}} & \frac{1}{\sqrt{(1 - \sqrt{3})^2 + 2}} \\ 
\frac{\sqrt{3} + 1}{\sqrt{(\sqrt{3} + 1)^2 + 2}} & \frac{1}{\sqrt{(\sqrt{3} + 1)^2 + 2}} & \frac{1}{\sqrt{(\sqrt{3} + 1)^2 + 2}}  
\end{array}
\right)
\end{equation}
The symmetry matrix $T$ is obtained by transforming $T_D$ to the $U$ basis. $T$ has the form
\begin{widetext}
\begin{equation}\label{eq:Symmat}
T=
\begin{pmatrix}
\frac{1}{6} \left( - (3 - \sqrt{3}) e^{i\theta_2} + (3 + \sqrt{3}) e^{i\theta_3} \right) &
\frac{-e^{i\theta_2} + e^{i\theta_3}}{2\sqrt{3}} & 
\frac{-e^{i\theta_2} + e^{i\theta_3}}{2\sqrt{3}} \\

\frac{-e^{i\theta_2} + e^{i\theta_3}}{2\sqrt{3}} & 
\frac{(3+\sqrt{3})e^{i\theta_1} + (2+\sqrt{3})e^{i\theta_2} + e^{i\theta_3}}{2(3+\sqrt{3})} & 
\frac{-(3+\sqrt{3})e^{i\theta_1} + (2+\sqrt{3})e^{i\theta_2} + e^{i\theta_3}}{2(3+\sqrt{3})} \\

\frac{-e^{i\theta_2} + e^{i\theta_3}}{2\sqrt{3}} & 
\frac{-(3+\sqrt{3})e^{i\theta_1} + (2+\sqrt{3})e^{i\theta_2} + e^{i\theta_3}}{2(3+\sqrt{3})} & 
\frac{(3+\sqrt{3})e^{i\theta_1} + (2+\sqrt{3})e^{i\theta_2} + e^{i\theta_3}}{2(3+\sqrt{3})}.
\end{pmatrix}
\end{equation}
\end{widetext}
The matrix in Eq.~(\ref{eq:Symmat}) leaves the Walsh-Hadamard matrix and its Hamiltonian invariant regardless of the values of $\theta_1$, $\theta_2$, and $\theta_3$. As an example, assigning $\theta_1=\pi$, $\theta_2 = 0$, and $\theta_3=0$, the matrix $S$ from  Eq.~(\ref{eq1:smat}) is obtained.  
By choosing the values to be zero, the resulting matrix will be the identity matrix, as expected.  
Applying the symmetry matrix $T$ from Eq.~(\ref{eq:Symmat}) to $U_d$ and $U_o$ will change them in such a way that their product will still be the Walsh-Hadamard matrix:
\begin{equation}\label{eq:sim}
    TWT^{-1} = TU_d U_oT^{-1} = T U_d T^{-1} T U_o T^{-1} = U_d^\prime U_o^{\prime} =  W
\end{equation}
since $TT^{-1} = T^{-1}T= \mathbbm{1}$. $U_d^\prime$ and $U_o^\prime$ are not necessarily the same as $U_d$ and $U_o$ ($U_d\neq U_d^\prime$ and $U_o\neq U_o^\prime$).  However,  $U_dU_o=U_d^\prime U_o^\prime=W$.   This shows how the other sets of parameters in Table \ref{tbl:tblS1} are obtained.  More details and another example are in given in Appendix \ref{app:ms}. 


\section{Relation between different sets of parameters for Permutations}

Now, let's look at different unitaries with more symmetries than a random unitary, as shown in FIG. \ref{fig:simu}. The first matrix is $P_1$
\begin{equation}
    P_1 = \left(\begin{array}{ccc}
     0 & 0 & e^{-i\xi_1} \\
     e^{-i\xi_2} & 0 & 0 \\
     0 & e^{-i\xi_3} & 0 
    \end{array}
    \right)
\end{equation}
which can also be written as
\begin{eqnarray}
    P_1 
    = 
    \left(\begin{array}{ccc}
     e^{-i\xi_1} & 0 & 0 \\
     0 & e^{-i\xi_2} & 0 \\
     0 & 0 & e^{-i\xi_3} 
    \end{array}
    \right)
    \left(\begin{array}{ccc}
     0 & 0 & 1 \\
     1 & 0 & 0 \\
     0 & 1 & 0 
    \end{array}
    \right).
\end{eqnarray}

 As a simplified case, let's assume $\xi_1=\xi_2=\xi_3=0$. In that case, the eigenvector and the symmetry matrices for $P_1$ are
\begin{equation*}
D_{P_1}=\frac{1}{\sqrt{3}}\left (
  \begin {array} {ccc}  1 &\frac {\sqrt {3}i - 
                1 } {2}   & - \frac {\sqrt {3}i + 
              1 } {2}    \\  1 & - \frac {\sqrt {3}i \
+1 } {2}   &\frac {\sqrt {3}i -1 
         } {2}    \\  1 &  1 &  1  \\\end {array}  \right)
\end{equation*}
\begin{widetext}
\begin{eqnarray*}
T_{P_1}
=\frac {1} {3}\left (
  \begin {array} {ccc}e^{-i  \gamma_1} + e^{-i  \gamma_2} + 
                    e^{-i  \gamma_3}  &e^{-i  \gamma_1} 
- e^{i\left(\frac{2\pi}{3}-\gamma_2\right)}   +e^{i\left(\frac{2\pi}{3}-\gamma_3\right)}    &e^{-i 
 \gamma_1} + e^{i\left(\frac{2\pi}{3}-\gamma_2\right)} - e^{i\left(\frac{2\pi}{3}-\gamma_3\right)}     \\e^{-i  \gamma_1} + e^{i\left(\frac{2\pi}{3}-\gamma_2\right)} - e^{i\left(\frac{2\pi}{3}-\gamma_3\right)}  &e^{-i  \gamma_1} + e^{-i  \gamma_2} +
              e^{-i  \gamma_3}  &e^{-i  \gamma_1} - 
e^{i\left(\frac{2\pi}{3}-\gamma_2\right)}   + e^{i\left(\frac{2\pi}{3}-\gamma_3\right)}  \\e^{-i  
\gamma_1} - e^{i\left(\frac{2\pi}{3}-\gamma_2\right)}   + e^{i\left(\frac{2\pi}{3}-\gamma_3\right)}  &e^{-i  \gamma_1} 
+ e^{i\left(\frac{2\pi}{3}-\gamma_2\right)} - e^{i\left(\frac{2\pi}{3}-\gamma_3\right)}    &e^{-i  
\gamma_1} + e^{-i  \gamma_2} + e^{-i  \gamma_3}   \\\end {array}  \right),
\end{eqnarray*}
\end{widetext}
if we choose 
$$T_{D_1}=\left(\begin{array}{ccc}
  e^{-i\gamma_1} & 0 & 0 \\
  0 & e^{-i\gamma_2} & 0 \\
  0 & 0 & e^{-i\gamma_3}
\end{array}\right).$$
Choosing values of phases for $T_{P_1}$ as $\gamma_1=0$, $\gamma_2=\frac{2\pi}{3}$, and $\gamma_3=\frac{4\pi}{3}$, the matrix $T_{P_1}(\xi_1=\xi_2=\xi_3=0)=S^\prime$ will be,
$$S^\prime=\left(\begin{array}{ccc}
     0 & 0 & 1 \\
     1 & 0 & 0 \\
     0 & 1 & 0 
    \end{array}
    \right)$$
 as a symmetry matrix, which won't affect the matrix $P_1$. As an example, in the table. \ref{table:p1}, looking at sets 4 and 8, the connection between them is shown as $$U_{d4}U_{o4}=S^\prime U_{d4}S^{\prime-1}S^\prime U_{o4}S^{\prime-1}= U_{d8}U_{o8},$$
 using $S^\prime$. The same transformation is evident between sets 5 and 7.
 
 Finding these connections requires following steps which are similar to those used in the case of the WH matrix. First, we start by looking at the diagonal Gell-Mann matrices, $\lambda_3, \lambda_8$, and $I$: 
 $$
TU_dT^\dagger = U_d^\prime \Rightarrow TH_dT^\dagger = H_d^\prime,
$$
$$
H_d = \sum_i a_i \lambda_i = a_0 I + \sum_{i=3,8} a_i \lambda_i,
$$
where $H_d$ represents the phases for one set of parameters and $H_d^\prime$ represents the phases for another set. The symmetry transformation for these three Gell-Mann matrices, using the symmetry matrix,  will assign values to the parameters defined in the symmetry matrix. In the case of $P_1$, these parameters are shown by $\gamma_1$, $\gamma_2$, and $\gamma_3$. Then we can use the same symmetry transformation by implementing the resulted symmetry matrix with the assigned values to find the off-diagonal elements of the off-diagonal part of the decomposition process in Sec.\ref{sec:WH}
$$
TU_oT^\dagger = U_o^\prime \Rightarrow TH_oT^\dagger = H_o^\prime.
$$
$$H_o=\sum_{\substack{j=1, \\ j\neq 3}}^7a_j\lambda_j.$$
Finally, we can look at the product of the two matrices after using the symmetry matrix on one set of parameters to get another set of parameters
 \begin{equation}
TUT^\dagger  = TU_dT^\dagger TU_o T^\dagger = U^\prime_d U^\prime_o.
\end{equation}

In other words, first, the symmetries of the matrix are identified by diagonalizing the gate, $U$, or equivalently, its Hamiltonian, $H_g$.  This defines a basis for the diagonal matrices that will commute with that gate.  $T_D$ is the matrix that is composed of a matrix that is diagonal in the same basis as $U$, but has been transformed to the $U$ basis.  In other words, if $H \rightarrow H_D$, where $H$ Hamiltonian that generates $U$ and $H_D$ is the set of eigenvalues of $H$, the diagonalized $H$, then the basis for $U$ is defined by the matrix $V$ where $VT_dV^\dagger = T,$ and $VH_dV^\dagger =H$.  Then, the effects of the symmetry matrix on the factors, $U_d$ and $U_o$ are determined and the possible variations of $U_d$ and $U_o$.  

 The next matrix, $P_2$, is a permutation between the first and second excited states of a qutrit and keeps the ground state in its original state.
\begin{equation}
    P_2 = \left(\begin{array}{ccc}
     1 & 0 & 0 \\
     0 & 0 & 1 \\
     0 & 1 & 0 
    \end{array}
    \right).
\end{equation}

We can do the same thing for $P_2$, where
\begin{equation*}
   D_{P_2} = \left (
  \begin {array} {ccc}  0 &  0 &  1  \\ - \frac {1} {\sqrt {2}} 
&\frac {1} {\sqrt {2}} &  0  \\\frac {1} {\sqrt {2}} &\frac {1} 
{\sqrt {2}} &  0  \\\end {array}  \right)
\end{equation*}
and

    \begin{equation*}
        T_{P_2}=\left (
  \begin {array} {ccc}  e^{-i  \phi_3} &  0 &  0  \\  0 &\frac {e^{-i  \phi_1} + 
              e^{-i  \phi_2} } 
{2} &\frac {e^{-i  \phi_2} - 
           e^{-i  \phi_1} } {2}  \\  0 &\frac {e^{-i  \phi_2} - 
        e^{-i  \phi_1} } {2} &\frac {e^{-i  \phi_1} + 
     e^{-i  \phi_2} } {2}  \\\end {array}  \right),
    \end{equation*}
    for its corresponding diagonal matrix 
     $$T_{D_2}=\left(\begin{array}{ccc}
  e^{-i\phi_1} & 0 & 0 \\
  0 & e^{-i\phi_2} & 0 \\
  0 & 0 & e^{-i\phi_3}
\end{array}\right).$$

\begin{table*}[ht]
	\caption{Numerically determined matrix elements of the matrix $P_1$ in the diagonal-off-diagonal decomposition.}
	\label{table:p1}
	\centering
	\begin{tabular*}{7in}{@{\extracolsep{\fill} } c c c c c c c }
		\hline
		\hline
		Decomposition & $m_{01}$ & $m_{12}$ & $m_{02}$ &$\phi_0$&$\phi_1$&$\phi_2$ \\
		\hline
		1&$0.0397 - 1.2085 i$&$-0.1990 - 1.1927i$ &$0.1597 + 1.1986 i$& $-0.1325$ & $-0.0329$ & $0.1654$ \\
        \hline
        2&$ 0.6956 - 0.9890 i$&$ 0.1779 - 1.1960i$ &$-0.8336 + 0.8758i $& $0.7607$ & $5.6702$ & $6.1355$ \\
        \hline
        3&$ 0.8035 - 0.9035i$&$ 0.9031 - 0.8040i$ &$-1.2092 + 0.0006i $& $ 1.5703$ & $5.5563 $ & $5.4398 $ \\
        \hline
        4&$ -0.9690 + 0.7232i$&$ 0.8013 - 0.9055i$ &$1.2050 + 0.1004i $& $ 2.3978$ & $4.2476 $ & $5.9210 $ \\
        \hline
        5&$ 0.5656 + 1.0701i$&$ 0.6404 + 1.0256i$ &$1.0454 + 0.6162i $& $ 2.6284$ & $4.9465 $ & $4.9915 $ \\
        \hline
        6&$ 1.2091 + 0.0137i$&$ 1.1582 + 0.3792i$ &$0.3922 - 1.1438i $& $ 3.4720$ & $ 4.7010$ & $ 4.3934$ \\
        \hline
        7&$ 1.0547 - 0.5913i$&$ 0.5601 + 1.0195i$ &$0.5714 - 1.0656i $& $ 4.9585$ & $2.6117 $ & $4.9662 $ \\
        \hline
        8&$ 1.1784 - 0.2079i$&$ -0.9690 + 0.7232i$ &$0.7346 + 0.9604i $& $ 5.9567$ & $ 2.2432$ & $ 4.3665$ \\
        \hline
        9&$ -1.1051 - 0.4907i$&$ 1.0882 - 0.5271i$ &$0.04013 + 1.2085i $& $6.2501 $ & $ 1.1529$ & $ 5.1635$ \\
        \hline
        10&$-0.3246 + 1.1648 i$&$ 0.4189 + 1.1343i$ &$0.0990 + 1.2051i $& $ 6.2012 $ & $ 2.8698$ & $ 3.4954$ \\
        \hline
        11&$ 0.5713 + 1.0657i$&$ -0.4813 + 1.1092i$ &$0.0998 + 1.2050i $& $ 6.2005  $ & $ 3.6337$ & $ 2.7322$ \\
         \hline
        12&$ 1.1726 - 0.2950i$&$ -1.1947 - 0.1865i$ &$0.1106 + 1.2041i $& $ 6.1916   $ & $ 4.9589$ & $ 1.4159$ \\
        \hline
		\hline
	\end{tabular*}
\end{table*}
\begin{table*}[ht]
	\caption{Numerically determined matrix elements of the matrix $P_2$ in the diagonal-off-diagonal decomposition.}
	\label{table:p2}
	\centering
	\begin{tabular*}{7in}{@{\extracolsep{\fill} } c c c c c c c }
		\hline
		\hline
		Decomposition & $m_{01}$ & $m_{12}$ & $m_{02}$ &$\phi_0$&$\phi_1$&$\phi_2$ \\
		\hline
        1&$ 0$&$ -0.5318 + 1.4780i $ &$ 0$& $0$ & $ 0.3454$ & $2.7962 $ \\
        \hline
        2&$ 0$&$ -1.2566 + 0.9424i $ &$0 $& $ 0$ & $0.9273 $ & $ 2.2143$ \\
        \hline
        3&$ 0$&$ -1.5170 + 0.4073i $ &$ 0$& $ 0$ & $ 1.3085$ & $ 1.8331$ \\
        \hline
		4&$ 0$&$-1.5558 - 0.2163 i $ &$ 0$& $0$ & $1.7090$ & $1.4326$ \\
        \hline
		5&$ 0$&$ 0.4587 + 1.5023 i$ &$0 $& $0$ & $2.2080$ & $0.9336$ \\
        \hline
        6&$ 0$&$ -0.3735 - 1.5257i $ &$ 0$& $ 0$ & $ 2.9015$ & $ 0.2401$ \\
        \hline
		7&$0 $&$ -1.5114 - 0.4277i$ &$ 0$& $0$ & $3.2795$ & $6.1453$ \\
        \hline
		8&$ 0$&$ 1.2607 - 0.9369i$ &$0 $& $0$ & $4.
        0733$ & $5.3515$ \\
        \hline
		9&$0 $&$1.2626 + 0.9344i $ &$0 $& $0$ & $5.3495$ & $4.0753$ \\
        \hline
         
        10&$0 $&$-1.4955 + 0.4803i $ &$0 $& $0$ & $6.1278$ & $3.2970$ \\
		\hline 
		\hline
	\end{tabular*}
\end{table*}
 Now, in Table \ref{table:p2}, sets 8 and 9 are related when swapping $\phi_1$ with $\phi_2$ and conjugating $m_{12}$. Here, $m_{01}$ and $m_{02}$ are equal to 0.  Notice that the same behavior happened in Table \ref{tbl:tblS1} between sets 3 and 4. With a slight deviation, the same pattern repeats between sets 7 and 10. This means that 
$$U_dU_o=SU_dS^{-1}SU_oS^{-1},$$ between sets 8 and 9 and sets 7 and 10. The symmetry matrix S was introduced in Eq. \ref{eq1:smat}, and when $\phi_2=\phi_3=0$, and $\phi_1=\pi$, $T_{P_2}(\phi_1=\pi,\phi_2=\phi_3=0)=S$.

\section{\label{sec:opt}The optimality of the possible evolutions}

 It is important to determine which parametrization is the most efficient by determining the optimal path when faced with several choices.  The optimal path may be determined by the shortest time, for example, or simply the shortest path.  Note that the two can be related by scaling the energy for the applied pulse.  
 The parametrization mentioned here is the diagonal-off-diagonal parametrization, which was introduced and experimentally motivated in \cite{PhysRevLett.125.180504}.
 Understanding this can guide us in optimizing our controls to reach the destination state, or unitary in the shortest time. 
 It is important for shortest-path determination that the diagonal part can be implemented by shifting the phase of the resonant pulses. In other words, applying the diagonal matrix $U_d$ involves a negligible experimental cost. This motivates minimizing the off-diagonal elements to obtain the shortest path.  

In \cite{PhysRevLett.125.180504} the decomposition chosen for the experiment was the one with the smallest coefficients of $\lambda_4$ and $\lambda_5$, more precisely the smallest combination $\alpha_4^2 + \alpha_5^2$. The reason for this choice is that the transition between states $\ket{0}$ and $\ket{2}$ was driven using a two-photon process, which requires a much higher drive power than the other two transitions. If we want to compare the different decompositions under the condition of having the same gate time and control pulse profile, we find that the required drive power is proportional to $\sqrt{\alpha_4^2 + \alpha_5^2}$. To minimize the drive power used to implement the gate, and hence minimize energy-level shifts and other side effects of strong driving, the decomposition that has the smallest value of $\alpha_4^2 + \alpha_5^2$ was chosen. It is possible that different physical implementations of qutrits can have different physical constraints than the superconducting qutrit in \cite{PhysRevLett.125.180504}. It is therefore possible that, in different scenarios, different sets of parameters would be optimal.  

In this context, consider the situation where the maximum Rabi frequency (maximum $\alpha_i$) for one of the three transitions in the qutrit is a fraction $\gamma$ of the maximum Rabi frequencies of the two other transitions. As in the case of the WH gate, the Rabi frequencies needed to implement the diagonal-off-diagonal decomposition of a generic qutrit unitary operator will all be comparable to each other. As a result, the following two scenarios can be compared: (1) implementing a single-pulse diagonal-off-diagonal decomposition with Rabi frequencies set by the smallest maximum Rabi frequency of the three transitions and (2) implementing a three-pulse Givens-rotation decomposition \cite{Givens1958ComputationOP, Frerix2019ApproximatingOM}, avoiding the slow transition and using the two transitions in the qutrit. Then, roughly speaking, the single-pulse approach is faster when $\gamma>1/3$, while the three-pulse approach is faster when $\gamma<1/3$. In fact, since the experiment of \cite{PhysRevLett.125.180504} used a flux qubit biased at the symmetry point where the maximum Rabi frequency of the two-photon transition is significantly smaller than the maximum Rabi frequencies of the single-photon transitions, the experiment belongs to the latter category, i.e.~the situation where a three-pulse Givens-rotation decomposition would result in a faster implementation of general qutrit gates.

Another possibility in the case where one of the transitions cannot be driven directly is a two-pulse sequence: a pulse with both allowed transitions driven resonantly preceded or followed by a pulse where only one of the two transitions is driven. It should also be possible to have two pulses where both transitions are driven in each pulse. This latter option has more adjustable parameters in the pulses than is necessary to reach an arbitrary qutrit unitary operator, which creates a situation where a continuum of solutions is expected to exist. Since this two-pulse protocol does not represent a fundamentally different approach to qutrit control and does not offer a dramatic speedup in experimental implementation, it will not be analyzed in detail in this paper.
 
In order to compare these to the fastest possible transition, without considering experimental constraints, an optimal path in parameter space can be compared to these implementations.  Using the Gell-Mann basis, the Walsh-Hadamard gate can be written as 
$e^{-i\hat{n}.\Vec{\lambda}\theta}$, where $\hat{n}$ is a unit vector, $
\theta$ is a parameter, and $\mbox{Tr}(\lambda_i\lambda_j) = 2\delta_{ij}$, for $i,j=0,1,\cdots,8$. To find the vector $\vec{n}$ the following method can be used.  Take the natural log of the Walsh-Hadamard matrix.  This is done by calculating $W=V W_d V^\dagger$, where $W_d$ is diagonal (or the eigenvalues and V is the eigenvector).  $\ln(W)=V(\ln W_d)V^\dagger =\vec{n}\cdot\vec{\lambda}.$  Then write $W(\theta) = e^{-i\hat{n}.\Vec{\lambda}\theta}$, so that $W(0)=\mathbbm{1}$, and $W(\theta)=W$ for $\theta = |\vec{n}|$.  In this case, $\theta = \frac{\pi}{2}$ including the $\lambda_0$ matrix to get the unitary matrix.  One can then compare this form to  $e^{-iG_d}e^{-iG_o}=e^{-i(\Vec{n}_1.\Vec{\lambda_1})t_1}e^{-i(\Vec{n}_2.\Vec{\lambda_2})t_2}$, Where 
$\Vec{\lambda_i}=\{\lambda_0,\lambda_3,\lambda_8\}$ and $\Vec{\lambda_j}=\{\lambda_1,\lambda_2,\lambda_4,\lambda_5,\lambda_6,\lambda_7\}$.  
Using the Hamiltonian
\begin{equation}
    H_W = \left(\begin{array}{ccc}
        -1+\frac{1}{\sqrt{3}} & \frac{1}{\sqrt{3}} & \frac{1}{\sqrt{3}} \\
        \frac{1}{\sqrt{3}} & -1-\frac{1}{2\sqrt{3}} & -\frac{1}{2\sqrt{3}} \\
        \frac{1}{\sqrt{3}} & -\frac{1}{2\sqrt{3}} & -1-\frac{1}{2\sqrt{3}} 
    \end{array}\right)
\end{equation}
will result in the Walsh-Hadamard matrix by exponentiation
\begin{equation}
    W = \exp[-iH_W(\pi/2)].
\end{equation}

These Hamiltonians are compared in Table \ref{tbl:opt}.  The first five sets are derived from Table \ref{tbl:tblS1}. The last set was calculated using $e^{-i(\vec{n}\cdot\vec{\lambda})t}$.
\begin{table*}[ht]
	\caption{Comparing the strength of the Hamiltonian of the 5 different sets of parametrization from Table \ref{tbl:tblS1} and the parametrization using the Gell-Mann basis, $e^{-i\vec{n}\cdot \vec{\lambda}}$.}
	\label{tbl:opt}
	\centering
	\begin{tabular*}{7in}{@{\extracolsep{\fill} } c c c c c c c }
		\hline
		\hline
		Decomposition & $(\vec{n_1}.\vec{n_1})t_1^2$ & $(\vec{n_2}.\vec{n_2})t_2^2$  & $(\vec{n_1}.\vec{n_1})t_1^2+(\vec{n_2}.\vec{n_2})t_2^2$ &$(\vec{n}.\vec{n})t^2$ \\
		\hline
		1&$0.4878$&$5.7251$ &$6.2129$& - 
        \\
        \hline
		2&$15.746$&$5.7248$ &$21.4708$& - \\
        \hline
		3&$ 15.7464$&$5.7248$ &$21.4708$&  - \\
        \hline
		4&$39.1583$&$5.8481$ &$45.0064$&   -\\
        \hline
    5&$34.7688$&$1.462$ &$36.2308$&  -\\
    \hline
    6&-& -&-& $4.1123$ \\
		\hline 
		\hline
	\end{tabular*}
\end{table*}

Clearly, the last set is the smallest and the fourth set has the largest value.   
The shortest path in parameter space from the identity matrix to Walsh-Hadamard can be calculated using the sum of the squares of the parameters and is described by the parametrization of $\vec{n}\cdot\vec{\lambda}$.  

In relation to the experimental implementation of the diagonal-off-diagonal decomposition, it is important to have the experimental ability to precisely perform qutrit operations with arbitrary Gell-Mann matrix coefficients. For this purpose, the different transitions can be calibrated individually and then combined with the necessary proportions to obtain the generator for any desired qutrit gate. Indeed, the authors of \cite{PhysRevLett.125.180504} followed this approach of calibrating the transition dynamics individually and then used this calibration data to construct the WH gate generator. One complication in that experiment was the relatively large energy shifts caused by the strong driving of the two-photon transition. However, with the proper calibration, the authors of \cite{PhysRevLett.125.180504} were able to account for and cancel these shifts.

\section{Conclusion}

Qutrits can have some advantages over qubits in quantum information processing \cite{Goss_2022,  PhysRevA.67.062313, PhysRevA.101.022304, 10.3389/fphy.2020.589504, 2009NatPh.5.134L}. However, the control of these systems is not as well understood in many devices.  The parametrization $U_oU_d$, which was introduced in \cite{PhysRevLett.125.180504} for superconducting qutrits was investigated.  This parametrization was shown to produce different parameter sets that achieve the same overall unitary evolution. For this parametrization, the following questions were answered.  Why are there different parameter sets that give the same unitary matrix?  How are these sets related?  Which sets should be used for optimal performance in a typical experimental scenario? These questions were answered using symmetry arguments.  When the symmetry of the matrix is not the symmetry of the constituent parts, a new set of parameters arises where the $U_o$ and the $U_d$ parts can compensate for each other.  This general method was used to explain each of the different parameter sets in \cite{PhysRevLett.125.180504}.  The example of the Walsh-Hadamard matrix, useful for Fourier transform operations, was experimentally implemented in \cite{PhysRevLett.125.180504}.  The results of that paper were described in detail including the origin and relationship among different sets of parameter values.  

In addition, an argument was provided for finding the optimal path in parameter space, and time-optimal control was discussed.  These arguments are general and apply to the $U_oU_d$ parametrization as well as the experimentally motivated Cartan decompositions listed in the appendix.  These are possible alternatives to the $U_oU_d$ decomposition.  In these cases, two-photon transitions can be replaced by single-photon transitions.

Note that the parametrizations provided are not the only possible ones. Any sets satisfying Eq.~(\ref{eq:cdc}) can be used to define another decomposition.  The ones presented here are experimentally motivated for systems that have an energy level structure similar to the one described in Ref.~\cite{PhysRevLett.125.180504}. 

Likewise, the argument used for the symmetries in the decomposition $U=U_dU_o$ that the overall symmetry  (of $U$) is not necessarily the symmetry of the individual terms, $U_o$ and $U_d$.  This is also true for gates that are not native. Furthermore, for larger systems consisting of several qutrits, the same argument holds, the symmetry of the overall evolution may not be the symmetry of the individual gates.  This topic will be discussed elsewhere.

\section*{Acknowledgment}
SA was supported by Japan's Ministry of Education, Culture, Sports, Science and Technology (MEXT) Quantum Leap Flagship Program Grant Number JPMXS0120319794.  MSB acknowledges the support of the National Science Foundation's IR/D program and Southern Illinois University.  The opinions and conclusions expressed herein are those of the authors, and do not represent any funding agencies.
LAW was supported by the grant PID2021-126273NB-I00 funded by MCIN/AEI/10.13039/501100011033, and by “ERDF A way of making Europe” and the Basque Government through Grant No. IT1470-22. This work has been financially sup- ported by the Ministry for Digital Transformation and of Civil Service of the Spanish Government through the QUAN- TUM ENIA project call – Quantum Spain project, and by the European Union through the Recovery, Transformation and Resilience Plan – NextGenerationEU within the frame- work of the Digital Spain 2026 Agenda. 
\section*{DATA AVAILABILITY}
The data that support the findings of this article are included in this paper.

\appendix
\section{Algebra of 3-state systems}\label{sec:AppendixA}

The Gell-Mann basis is:
\begin{eqnarray}\label{gellmann}
\lambda_1 = \left(\begin{array}{ccc}
		0 & 1 & 0 \\
		1 & 0 & 0 \\
		0 & 0 & 0 
			\end{array}\right), && \;\;
\lambda_2 = \left(\begin{array}{ccc}
		0 & -i & 0 \\
		i & 0 & 0 \\
		0 & 0 & 0 
			\end{array}\right), \nonumber \\
\lambda_4 = \left(\begin{array}{ccc}
		0 & 0 & 1 \\
		0 & 0 & 0 \\
		1 & 0 & 0 
			\end{array}\right), && \;\;
\lambda_5 = \left(\begin{array}{ccc}
		0 & 0 & -i \\
		0 & 0 & 0 \\
		i & 0 & 0 
			\end{array}\right), \nonumber \\
\lambda_6 = \left(\begin{array}{ccc}
		0 & 0 & 0 \\
		0 & 0 & 1 \\
		0 & 1 & 0 
			\end{array}\right), && \;\;
\lambda_7 = \left(\begin{array}{ccc}
		0 & 0 & 0 \\
		0 & 0 & -i \\
		0 & i & 0 
			\end{array}\right), \nonumber \\			
\lambda_3 = \left(\begin{array}{ccc}
		1 & 0 & 0 \\
		0 & -1 & 0 \\
		0 & 0 & 0 
			\end{array}\right),&& \;\;
\lambda_8 = \frac{1}{\sqrt{3}}
	    \left(\begin{array}{ccc}
		1 & 0 & 0 \\
		0 & 1 & 0 \\
		0 & 0 & -2 \end{array}\right),\nonumber\\
        \lambda_0 = \sqrt{\frac{2}{3}}\left(\begin{array}{ccc}
		1 & 0 & 0 \\
		0 & 1& 0 \\
		0 & 0 & 1 
			\end{array}\right).
\end{eqnarray}

For this algebra, the commutation and anticommutation relations are given by 
\begin{equation}
    \lambda_i\lambda_j = \frac{2}{3}I \delta_{ij} +if_{ijk}\lambda_k +d_{ijk}\lambda_k,
\end{equation}
with $[\lambda_i,\lambda_j] = 2if_{ijk}\lambda_k$ and $\{\lambda_i,\lambda_j\} = \frac{4}{3}I\delta_{ij}+2d_{ijk}\lambda_k$ \cite{ampv},
where $I$ is the $3\times 3$ identity matrix and the sum over repeated indices is understood.  

Explicitly, for the Gell-Mann basis, the nonzero symmetric $d_{ijk}$ are
\begin{eqnarray}
d_{118}\!&=&\! d_{228}= d_{338}= -d_{888}= \frac{1}{\sqrt{3}}, \nonumber\\
d_{448}\!&=& \!d_{558}= d_{668}= d_{778}=-\frac{1}{2\sqrt{3}}, \nonumber\\ 
d_{146}\!&=&\! d_{157}= -d_{247}= d_{256} = d_{344} \nonumber \\
\!&=&\! d_{355}= -d_{366}= -d_{377}=\frac{1}{2}.
\end{eqnarray}

The nonzero anti-symmetric $f_{ijk}$ are
\begin{gather}\label{eq:su(3)fs}
f_{147}= f_{246}= f_{257}= f_{345}=-f_{156}=-f_{367}=\frac{1}{2}, \nonumber\\ 
f_{123}= 1, \;\;\;\;\;\;\;\;\;\;\;\;\;
f_{458}= f_{678}=\frac{\sqrt{3}}{2}.
\end{gather}

Let us define several other diagonal matrices that can be used to complete a basis of traceless, Hermitian matrices.  For comparison, the ones used in the standard Gell-Mann basis are given first: 
\begin{eqnarray}
\lambda_3 &=& \left(\begin{array}{ccc}  1 & 0 & 0 \\ 0 & -1 & 0 \\ 0 & 0 & 0 \end{array}\right),  \\
\lambda_9 &=& 
\left(\begin{array}{ccc}  1 & 0 & 0 \\ 0 & 0 & 0 \\ 0 & 0 & -1 \end{array}\right), \;\;
\lambda_{10} = 
\left(\begin{array}{ccc}  0 & 0 & 0 \\ 0 & 1 & 0 \\ 0 & 0 & -1 \end{array}\label{eq:sec}\right).
\end{eqnarray}
This would determine the form of the other independent diagonal matrix.  It would be 
\begin{eqnarray}
\lambda_8 &=&\frac{1}{\sqrt{3}}\left(\begin{array}{ccc}  
1 & 0 & 0 \\ 
0 & 1 & 0 \\ 
0 & 0 & -2 \end{array}\right), \\
\lambda_{11} &=& 
\frac{1}{\sqrt{3}}\left(\begin{array}{ccc}  
1 & 0 & 0 \\ 
0 & -2 & 0 \\ 
0 & 0 & 1 \end{array}\right),
\lambda_{12} = 
\frac{1}{\sqrt{3}}\left(\begin{array}{ccc} 
-2 & 0 & 0 \\ 
0 & 1 & 0 \\ 
0 & 0 & 1 \end{array}\label{eq:thir}\right).
\end{eqnarray}
We can find a new set of symmetric and anti-symmetric constants using the matrices in Eq.~(\ref{eq:sec}) and Eq.~(\ref{eq:thir}) and their
commutation and anti-commutation relations.
\begin{table}
         \setlength\tabcolsep{0.05pt}
        \begin{tabular}{|c|c|c|c|c|c|c|c|c|}
        \hline
            &$\boldsymbol{\lambda_1}$ &  $\boldsymbol{\lambda_2}$ &  $\boldsymbol{\lambda_{3}}$ &  $\boldsymbol{\lambda_4}$  &  $\boldsymbol{\lambda_5}$ &  $\boldsymbol{\lambda_6}$ &  $\boldsymbol{\lambda_7}$ &  $\boldsymbol{\lambda_{8}}$   \\
            \hline
            $\boldsymbol{\lambda_{1}}$  &0 &  $2i\lambda_{3}$& ${\raisebox{1pt}{\text{-}}}2i\lambda_{2}$& $i\lambda_{7}$&  ${\raisebox{1pt}{\text{-}}}i\lambda_{6}$& $i\lambda_{5}$& ${\raisebox{1pt}{\text{-}}}i\lambda_{4}$ &  $0$ 
            \\
           \hline
           $\boldsymbol{\lambda_{2}}$  & ${\raisebox{1pt}{\text{-}}}2i\lambda_{3}$ &  0& $2i\lambda_{1}$& $i\lambda_{6}$&  $i\lambda_{7}$& ${\raisebox{1pt}{\text{-}}}i\lambda_{4}$&  ${\raisebox{1pt}{\text{-}}}i\lambda_{5}$&   $0$ \\
           \hline
            $\boldsymbol{\lambda_{3}}$  &  $2i\lambda_{2}$&   ${\raisebox{1pt}{\text{-}}}2i\lambda_{1}$& 0& $i\lambda_{5}$&   ${\raisebox{1pt}{\text{-}}}i\lambda_{4}$&  ${\raisebox{1pt}{\text{-}}}i\lambda_{7}$&  $i\lambda_{6}$ &   0 \\
             \hline
            $\boldsymbol{\lambda_{4}}$  & ${\raisebox{1pt}{\text{-}}}i\lambda_{7}$&  ${\raisebox{1pt}{\text{-}}}i\lambda_{6}$& ${\raisebox{1pt}{\text{-}}}i\lambda_{5}$& 0 & $i\lambda_{3}$ &$i\lambda_{2}$ &  $i\lambda_{1}$& ${\raisebox{1pt}{\text{-}}}i\sqrt{3}\lambda_{5}$    \\
            ~ & ~ &~& ~ & ~ & $+i\sqrt{3}\lambda_{8}$ & ~ & ~ &~ \\
             \hline
            $\boldsymbol{\lambda_{5}}$  & $i\lambda_{6}$&  ${\raisebox{1pt}{\text{-}}}i\lambda_{7}$& $i\lambda_{4}$& ${\raisebox{1pt}{\text{-}}}i\lambda_{3}$ & 0 &${\raisebox{1pt}{\text{-}}}i\lambda_{1}$ &  ${\raisebox{1pt}{\text{-}}}i\lambda_{2}$& $i\sqrt{3}\lambda_{4}$   \\
            ~ & ~& ~ & ~ & ${\raisebox{1pt}{\text{-}}}i\sqrt{3}\lambda_{8}$ & ~ & ~ & ~ &~ \\
             \hline
              $\boldsymbol{\lambda_{6}}$  & ${\raisebox{1pt}{\text{-}}}i\lambda_{5}$&  $i\lambda_{4}$& $i\lambda_{7}$& ${\raisebox{1pt}{\text{-}}}i\lambda_{2}$ & $i\lambda_{1}$ &0 &  ${\raisebox{1pt}{\text{-}}}i\lambda_{3}$& ${\raisebox{1pt}{\text{-}}}i\sqrt{3}\lambda_7$    \\
            ~ &  ~& ~& ~&  ~& ~& ~&  $+i\sqrt{3}\lambda_8$ &\\
             \hline
            $\boldsymbol{\lambda_{7}}$  & $i\lambda_{4}$&  $i\lambda_{5}$& ${\raisebox{1pt}{\text{-}}}i\lambda_{6}$& ${\raisebox{1pt}{\text{-}}}i\lambda_{1}$ & $i\lambda_{2}$ &$i\lambda_{3}$ &  0& $i\sqrt{3}\lambda_6$   \\
            ~ &  ~& ~& ~&  ~& ~& ${\raisebox{1pt}{\text{-}}}i\sqrt{3}\lambda_8$&  ~ &\\
             \hline
          $\boldsymbol{\lambda_{8}}$  & 0&  0& 0& $i\sqrt{3}\lambda_{5}$ & ${\raisebox{1pt}{\text{-}}}i\sqrt{3}\lambda_{4}$ &$i\sqrt{3}\lambda_{7}$ &  ${\raisebox{1pt}{\text{-}}}i\sqrt{3}\lambda_{6}$& 0    \\
             \hline
             
        \end{tabular}
         \caption{Commutation relations of the Gell-Mann basis using the matrices in Eq. (\ref{gellmann}).}
	\label{tbl:tbl1}
    \end{table}

 \begin{table}
     \setlength\tabcolsep{0.03pt}
        \begin{tabular}{|c|c|c|c|c|c|c|c|c|}
        \hline
            &$\boldsymbol{\lambda_1}$ &  $\boldsymbol{\lambda_2}$ &  $\boldsymbol{\lambda_9}$ &  $\boldsymbol{\lambda_4}$  &  $\boldsymbol{\lambda_5}$ &  $\boldsymbol{\lambda_6}$ &  $\boldsymbol{\lambda_7}$ &  $\boldsymbol{\lambda_{11}}$   \\
            \hline
            $\boldsymbol{\lambda_{1}}$  &0 &  $i\sqrt{3}\lambda_{11}$& ${\raisebox{1pt}{\text{-}}}i\lambda_{2}$& $i\lambda_{7}$&  ${\raisebox{1pt}{\text{-}}}i\lambda_{6}$& $i\lambda_{5}$& ${\raisebox{1pt}{\text{-}}}i\lambda_{4}$ &  ${\raisebox{1pt}{\text{-}}}i\sqrt{3}\lambda_{2}$ 
            \\
            ~ &  ~& $+i\lambda_{9}$& ~&  ~& ~& ~ &  ~&
            \\
           \hline
           $\boldsymbol{\lambda_{2}}$  & ${\raisebox{1pt}{\text{-}}}i\lambda_{9}$ &  0& $i\lambda_{1}$& $i\lambda_{6}$&  $i\lambda_{7}$& ${\raisebox{1pt}{\text{-}}}i\lambda_{4}$&  ${\raisebox{1pt}{\text{-}}}i\lambda_{5}$&   $i\sqrt{3}\lambda_{1}$\\
            ~ & ${\raisebox{1pt}{\text{-}}}i\sqrt{3}\lambda_{11}$ &~& ~ & ~ & ~ & ~ & ~ &~ \\
           \hline
            $\boldsymbol{\lambda_{9}}$  &  $i\lambda_{2}$&   ${\raisebox{1pt}{\text{-}}}i\lambda_{1}$& 0& $2i\lambda_{5}$&   ${\raisebox{1pt}{\text{-}}}2i\lambda_{4}$&  $i\lambda_{7}$&  ${\raisebox{1pt}{\text{-}}}i\lambda_{6}$ &   0 \\
             \hline
            $\boldsymbol{\lambda_{4}}$  & ${\raisebox{1pt}{\text{-}}}i\lambda_{7}$&  ${\raisebox{1pt}{\text{-}}}i\lambda_{6}$& ${\raisebox{1pt}{\text{-}}}2i\lambda_{5}$& 0 & $2i\lambda_{9}$ &$i\lambda_{2}$ &  $i\lambda_{1}$& 0    \\
             \hline
            $\boldsymbol{\lambda_{5}}$  & $i\lambda_{6}$&  ${\raisebox{1pt}{\text{-}}}i\lambda_{7}$& $2i\lambda_{4}$& ${\raisebox{1pt}{\text{-}}}2i\lambda_{9}$ & 0 &${\raisebox{1pt}{\text{-}}}i\lambda_{1}$ &  $i\lambda_{2}$& 0   \\
             \hline
              $\boldsymbol{\lambda_{6}}$  & ${\raisebox{1pt}{\text{-}}}i\lambda_{5}$&  $i\lambda_{4}$& ${\raisebox{1pt}{\text{-}}}i\lambda_{7}$& ${\raisebox{1pt}{\text{-}}}i\lambda_{2}$ & $i\lambda_{1}$ &0 &  $i\lambda_{9}$& $i\sqrt{3}\lambda_{7}$    \\
              ~ & ~& ~ & ~ & ~ & ~ & ~ & ${\raisebox{1pt}{\text{-}}}i\sqrt{3}\lambda_{11}$ &~ \\
             \hline
            $\boldsymbol{\lambda_{7}}$  & $i\lambda_{4}$&  $i\lambda_{5}$& $i\lambda_{6}$& ${\raisebox{1pt}{\text{-}}}i\lambda_{1}$ & ${\raisebox{1pt}{\text{-}}}i\lambda_{2}$ &$i\sqrt{3}\lambda_{11}$ &  0& ${\raisebox{1pt}{\text{-}}}i\sqrt{3}\lambda_{6}$    \\
            ~ & ~& ~ & ~ & ~ & ~ & ${\raisebox{1pt}{\text{-}}}i\lambda_{9}$ & ~ &~ \\
             \hline
          $\boldsymbol{\lambda_{11}}$  & $i\sqrt{3}\lambda_{2}$&  ${\raisebox{1pt}{\text{-}}}i\sqrt{3}\lambda_{1}$& 0& 0 & 0 &${\raisebox{1pt}{\text{-}}}i\sqrt{3}\lambda_{7}$ &  $i\sqrt{3}\lambda_{6}$& 0    \\
             \hline
             
        \end{tabular}
         \caption{Commutation relations of the Gell-Mann basis using the matrices in Eq. (\ref{gellmann}). For the diagonal elements, $\lambda_9$ and $\lambda_{11}$ were used
from Eqs. (\ref{eq:sec}) and (\ref{eq:thir}).}
	\label{tbl:tbl2}
    \end{table}
     
\begin{table}[ht]
         \setlength\tabcolsep{0.05pt}
        \begin{tabular}{|c|c|c|c|c|c|c|c|c|c|}
        \hline
            &$\boldsymbol{\lambda_1}$ &  $\boldsymbol{\lambda_2}$ &  $\boldsymbol{\lambda_{10}}$ &  $\boldsymbol{\lambda_4}$  &  $\boldsymbol{\lambda_5}$ &  $\boldsymbol{\lambda_6}$ &  $\boldsymbol{\lambda_7}$ &  $\boldsymbol{\lambda_{12}}$   \\
            \hline
            $\boldsymbol{\lambda_{1}}$  &0 &  $i\sqrt{3}\lambda_{12}$& $i\lambda_{2}$& $i\lambda_{7}$&  ${\raisebox{1pt}{\text{-}}}i\lambda_{6}$& $i\lambda_{5}$& ${\raisebox{1pt}{\text{-}}}i\lambda_{4}$ &  $i\sqrt{3}\lambda_{2}$ 
            \\
            ~ &  ~& ${\raisebox{1pt}{\text{-}}}i\lambda_{10}$& ~&  ~& ~& ~ &  ~&
            \\
           \hline
           $\boldsymbol{\lambda_{2}}$  & $i\lambda_{10}$ &  0& ${\raisebox{1pt}{\text{-}}}i\lambda_{1}$& $i\lambda_{6}$&  $i\lambda_{7}$& ${\raisebox{1pt}{\text{-}}}i\lambda_{4}$&  ${\raisebox{1pt}{\text{-}}}i\lambda_{5}$&   ${\raisebox{1pt}{\text{-}}}i\sqrt{3}\lambda_{2}$\\
            ~ & ${\raisebox{1pt}{\text{-}}}i\sqrt{3}\lambda_{12}$ &~& ~ & ~ & ~ & ~ & ~ &~ \\
           \hline
            $\boldsymbol{\lambda_{10}}$  &  ${\raisebox{1pt}{\text{-}}}i\lambda_{2}$&   $i\lambda_{1}$& 0& $i\lambda_{5}$&   ${\raisebox{1pt}{\text{-}}}i\lambda_{4}$&  $2i\lambda_{7}$&  ${\raisebox{1pt}{\text{-}}}2i\lambda_{6}$ &   0 \\
             \hline
            $\boldsymbol{\lambda_{4}}$  & ${\raisebox{1pt}{\text{-}}}i\lambda_{7}$&  ${\raisebox{1pt}{\text{-}}}i\lambda_{6}$& 
            ${\raisebox{1pt}{\text{-}}}i\lambda_{5}$& 0 & $i\lambda_{10}$ &$i\lambda_{2}$ &  $i\lambda_{1}$& $i\sqrt{3}\lambda_{5}$    \\
            ~ & ~ &~& ~ & ~ & ${\raisebox{1pt}{\text{-}}}i\sqrt{3}\lambda_{12}$ & ~ & ~ &~ \\
             \hline
            $\boldsymbol{\lambda_{5}}$  & $i\lambda_{6}$&  ${\raisebox{1pt}{\text{-}}}i\lambda_{7}$& $i\lambda_{4}$& $i\sqrt{3}\lambda_{12}$ & 0 &${\raisebox{1pt}{\text{-}}}i\lambda_{1}$ &  $i\lambda_{2}$& ${\raisebox{1pt}{\text{-}}}i\sqrt{3}\lambda_{4}$   \\
            ~ & ~& ~ & ~ & ${\raisebox{1pt}{\text{-}}}i\lambda_{10}$ & ~ & ~ & ~ &~ \\
             \hline
              $\boldsymbol{\lambda_{6}}$  & ${\raisebox{1pt}{\text{-}}}i\lambda_{5}$&  $i\lambda_{4}$& ${\raisebox{1pt}{\text{-}}}2i\lambda_{7}$& ${\raisebox{1pt}{\text{-}}}i\lambda_{2}$ & $i\lambda_{1}$ &0 &  $2i\lambda_{10}$& 0    \\
             \hline
            $\boldsymbol{\lambda_{7}}$  & $i\lambda_{4}$&  $i\lambda_{5}$& $2i\lambda_{6}$& ${\raisebox{1pt}{\text{-}}}i\lambda_{1}$ & ${\raisebox{1pt}{\text{-}}}i\lambda_{2}$ &${\raisebox{1pt}{\text{-}}}2i\lambda_{10}$ &  0& 0   \\
             \hline
          $\boldsymbol{\lambda_{12}}$  & ${\raisebox{1pt}{\text{-}}}i\sqrt{3}\lambda_{2}$&  $i\sqrt{3}\lambda_{2}$& 0& ${\raisebox{1pt}{\text{-}}}i\sqrt{3}\lambda_{5}$ & $i\sqrt{3}\lambda_{4}$ &0 &  0& 0    \\
             \hline
             
        \end{tabular}
         \caption{Commutation relations of the Gell-Mann basis using the matrices in Eq. (\ref{gellmann}). For the diagonal elements, $\lambda_{10}$ and $\lambda_{12}$ were used
from Eqs. (\ref{eq:sec}) and (\ref{eq:thir}).}
	\label{tbl:tbl3}
    \end{table}
    
 The anti-symmetric $f_{ijk}$ can be found using $$f_{ijk}=\frac{1}{4i}\tr([\lambda_i,\lambda_j]\lambda_k),$$ and the symmetric $d_{ijk}$ using
$$d_{ijk}=\frac{1}{4}\tr(\{\lambda_i,\lambda_j\}\lambda_k).$$  
Using Tables \ref{tbl:tbl2} and \ref{tbl:tbl3} the constants are:
\begin{gather}\label{eq:su(3)fs_2}
f_{129}= f_{147}= f_{246}= f_{257}=-f_{156}=f_{967}=\frac{1}{2}, \nonumber\\ 
f_{945}= 1, \;\;\;\;\;\;\;\;\;\;\;\;\;
f_{12,11}= -f_{67, 11}=\frac{\sqrt{3}}{2}.
\end{gather}

\begin{eqnarray}
d_{99,11}\!&=&\! d_{44,11}= d_{55,11}= -d_{11,11,11}= \frac{1}{\sqrt{3}}, \nonumber\\
d_{11,11}\!&=& \!d_{22,11}= d_{66,11}= d_{77,11}=-\frac{1}{2\sqrt{3}}, \nonumber\\ 
d_{119}\!&=&\! d_{146}= d_{157}= d_{229} = -d_{247} \nonumber \\
\!&=&\! d_{256}= -d_{966}= -d_{977}=\frac{1}{2}.
\end{eqnarray}

\begin{gather}\label{eq:su(3)fs_3}
-f_{12,10}= f_{147}= -f_{156}= f_{246}=f_{257}=f_{10,45}=\frac{1}{2}, \nonumber\\ 
f_{10,67}= 1, \;\;\;\;\;\;\;\;\;\;\;\;\;
f_{1,2,12}= f_{45, 12}=-\frac{\sqrt{3}}{2}.
\end{gather}

\begin{eqnarray}
d_{10,10,12}\!&=&\! d_{66,12}= d_{77,12}= -d_{12,12,12}= \frac{1}{\sqrt{3}}, \nonumber\\
d_{11,12}\!&=& \!d_{22,12}= d_{44,12}= d_{55,12}=-\frac{1}{2\sqrt{3}}, \nonumber\\ 
d_{11,10}\!&=&\! d_{146}= d_{157}= d_{22,10} = -d_{247} \nonumber \\
\!&=&\! d_{256}= -d_{10,44}= -d_{10,55}=\frac{1}{2}.
\end{eqnarray}


\section{Example:  How two sets of parameters are related}\label{app:ms}

Looking at table \ref{tbl:tblS1}, we want to show that these sets of parameters are related.
As an example, the relationship between two parameter sets (the second and the fifth set) is discussed in detail.  This provides an explicit example of how to obtain one set of parameters from another using the symmetry arguments in the text.

For the fifth set of parameters, using $S^2= \mathbbm{1}$, $U_5$ can be written as 
\begin{widetext}
\begin{eqnarray}\label{eq:u5dec}
U_5=&\left(\begin{array}{ccc} 
e^{-i\phi_0} & 0 & 0 
\\ 0 & e^{-i\phi_1}  & 0 
\\ 0 & 0 & e^{-i\phi_1}  \end{array}\right)
\left(\begin{array}{ccc}
   v_1 &  v_2 &  v_3 
\end{array}
\right)
{\left(\begin{array}{ccc}  e^{-ia} & 0 & 0 \\ 0 & e^{-ib}  & 0 \\ 
0 & 0 & e^{-ic}  \end{array}\right)}
\left(\begin{array}{ccc}  
   v_1^\ast   \\ 
 v_2^\ast   \\ 
  v_3^\ast  \end{array}\right)\\ \notag
=&\left(\begin{array}{ccc} 
e^{-i\phi_0} & 0 & 0 \\ 0 & e^{-i\phi_1}  & 0 \\ 0 & 0 & e^{-i\phi_1}  \end{array}\right)
\left(\begin{array}{ccc}
   v_1 &  v_2 &  v_3   
\end{array}\right) S \cdot S{\left(\begin{array}{ccc}  e^{-ia} & 0 & 0 \\ 0 & e^{-ib}  & 0 \\ 0 & 0 & e^{-ic}  \end{array}\right)}
S \cdot S
\left(\begin{array}{ccc}  
  v_1^\ast   \\ 
  v_2^\ast   \\ 
 v_3^\ast  \end{array}\right)\\ \notag
=&\left(\begin{array}{ccc} 
e^{-i\phi_0} & 0 & 0 \\ 0 & e^{-i\phi_1}  & 0 \\ 0 & 0 & e^{-i\phi_1}  \end{array}\right)
\left(\begin{array}{ccc}
   v_1 &  v_3 &  v_2 
\end{array}\right)
{\left(\begin{array}{ccc}  e^{-ia} & 0 & 0 \\ 0 & e^{-ic}  & 0 \\ 0 & 0 & e^{-ib}  \end{array}\right)}
\left(\begin{array}{ccc}  
  v_1^\ast  \\ 
 v_3^\ast   \\ 
  v_2^\ast   \end{array}\right).
\end{eqnarray}
\end{widetext}

The same decomposition can be used for the second set of parameters in the Table \ref{tbl:tblS1}.  Since every element of $G_o$ was multiplied by $-2$ the eigenvalues are also multiplied by $-2$ and the second and third eigenvalues will be exchanged by the action of $S$. 
To be more precise, the eigenvalues of the off-diagonal part after exponentiation will be changed, and the $-2$ factor will affect the exponents.  Also note that the second and third eigenvectors are exchanged; ($v_1$, $v_2$, $v_3$) $\rightarrow$ ($v_1$, $v_3$, $v_2$).
To find how the two sets 2 and 5 in Table \ref{tbl:tblS1} are related, note that $\phi_0$, $\phi_1$, a, b, c, $v_1$, $v_2$, $v_3$ are the same in $U_2$ and $U_5$. The idea here is to show that starting from $U_5$, $U_2$ can be obtained (or vice versa).  First decompose $U_2$ into its diagonal and off-diagonal parts as $U_{2d}U_{2o}$. Next decompose $U_{2o}$ into its eigenvalues and eigenvectors.  Then do the same for $U_5$. The first line in Eq.~(\ref{eq:u2dec}) is similar to the last line in Eq.~(\ref{eq:u5dec}). The differences between them are that first, the diagonal parts, $U_{2d}$ and $U_{5d}$, differ by $\pm\frac{2\pi}{3}$. The same difference was shown in Table \ref{tbl:tblS1} ($1.9199+\frac{4\pi}{3} = 6.1086$, $6.1086-\frac{2\pi}{3}=4.0143$). Since $\frac{-2\pi}{3} +2\pi = \frac{4\pi}{3}$, these two ($\frac{-2\pi}{3}$ and $\frac{4\pi}{3}$) are equivalent. Also note that the eigenvalue matrix of $U_{2o}$ has elements that are twice the fifth set (with a negative sign). The eigenvectors are the same for both of them.  
\begin{widetext}
\begin{eqnarray}\label{eq:u2dec}
U_2&=&\left(\begin{array}{ccc} 
e^{-i(\phi_0-\frac{4\pi}{3})} & 0 & 0 \\ 0 & e^{-i(\phi_1+\frac{2\pi}{3})}  & 0 \\ 0 & 0 & e^{-i(\phi_1+\frac{2\pi}{3})}  \end{array}\right)
\left(\begin{array}{ccc}
  v_1 &  v_3 &  v_2  
\end{array}\right)
\left(\begin{array}{ccc}  e^{-i(-2a)} & 0 & 0 \\ 0 & e^{-i(-2c)}  & 0 \\ 0 & 0 & e^{-i(-2b)}  \end{array}\right)
\left(\begin{array}{ccc}  
 v_1^\ast  \\ 
 v_3^\ast   \\ \notag
  v_2^\ast \end{array}\right)\\ \notag
  &=&\left(\begin{array}{ccc} 
e^{-i\phi_0} & 0 & 0 \\ 0 & e^{-i\phi_1}  & 0 \\ 0 & 0 & e^{-i\phi_1}  \end{array}\right)
\left(\begin{array}{ccc}
  v_1 &  v_3 &  v_2 
\end{array}\right) \left(\begin{array}{ccc}  e^{-i(-2a-\frac{4\pi}{3}+2\pi)} & 0 & 0 \\ 0 & e^{-i(-2c+\frac{2\pi}{3})}  & 0 \\ 0 & 0 & e^{-i(-2b+\frac{2\pi}{3})}  \end{array}\right)
\left(\begin{array}{ccc}  
 v_1^\ast  \\ 
 v_3^\ast   \\ 
  v_2^\ast\end{array}\right)\\ 
  &=&\left(\begin{array}{ccc} 
e^{-i\phi_0} & 0 & 0 \\ 0 & e^{-i\phi_1}  & 0 \\ 0 & 0 & e^{-i\phi_1}  \end{array}\right)
\left(\begin{array}{ccc}
  v_1 &  v_3 &  v_2 
\end{array}\right) \left(\begin{array}{ccc}  e^{-i(-2a-\frac{4\pi}{3})} & 0 & 0 \\ 0 & e^{-i(-2c+\frac{2\pi}{3})}  & 0 \\ 0 & 0 & e^{-i(-2b+\frac{2\pi}{3})}  \end{array}\right)
\left(\begin{array}{ccc}  
 v_1^\ast  \\ 
 v_3^\ast   \\ 
  v_2^\ast\end{array}\right).
\end{eqnarray}
\end{widetext}

Based on the table from \cite{PhysRevLett.125.180504}, the last line of $U_2$ and $U_5$, from Eq.~(\ref{eq:u2dec}) and Eq.~(\ref{eq:u5dec}), give the WH matrix.  If $U_5 = e^{-iG_{d5}}e^{-iG_{o5}}=e^{-iG_{d5}} V e^{-ig_d} V^\dagger$, then $U_2 = e^{-iG_{d2}}e^{-iG_{o2}}=e^{-iG_{d2}^\prime} V e^{-i(-2g_d + \frac{2\pi}{3})} V^\dagger$. Notice that $G_{d5}$ is equal to $G_{d2}^\prime$, and $$g_d = \left(\begin{array}{ccc}  a & 0 & 0 \\ 0 & c  & 0 \\ 0 & 0 & b \end{array}\right).$$

Therefore, the two linear equations can be solved using Eq.~(\ref{eq:u5dec}) and Eq.~( \ref{eq:u2dec}) since the last two lines are the same, except for the third matrix factor in each.  

This enables the determination of $a$ and $b$. These equations are $-2a-\frac{4\pi}{3} = a$ and $-2b+\frac{2\pi}{3}=b$ (the equations for $b$ and $c$ are the same). The results are $a = -1.3962$ and $b = c = 0.6981$.  


\section{\label{sec:car}Parametrizations using the Cartan Decomposition}
\label{sec:cd}

The parametrization $U_oU_d$ was not proven analytically, and it is clearly an overparametrization.  This is motivated by experimental considerations.  So here, an analytic form that is quite similar to this one is presented using the Cartan decomposition \cite{helgason1979differential, hermann1966lie}.    
The parametrization has the form
\begin{equation}
U = U_d U_{o1}U_{o2},
\end{equation}
where $U$ is any element of $SU(3)$ (or $U(3)$ as discussed), $U_d$ is a diagonal matrix, $U_{o1}$ and $U_{o2}$ are both off-diagonal.  The forms of $U_{o1}$ and $U_{o2}$ can be chosen in different ways, as will be shown.  

One form is 
\begin{eqnarray}\label{eq:cd}
G &=& e^{-i\left(\beta_3\lambda_3+\alpha_8\lambda_8  \right)}e^{-i\left(\beta_1\lambda_1+\beta_2\lambda_2  \right)} \nonumber \\
&&\times e^{-i\left(\alpha_4\lambda_4+\alpha_5\lambda_5+\alpha_6\lambda_6+\alpha_7\lambda_7  \right)},
\end{eqnarray}
where all $\alpha_i$ and $\beta_j$ are real parameters.  

Notice this is {\it not} of the form 
\begin{equation}
\label{eq:do}
G = e^{-i\left(\beta_3\lambda_3+\alpha_8\lambda_8  \right)} e^{-i\left(\alpha_1\lambda_1+\beta_2\lambda_2  + \alpha_4\lambda_4+\alpha_5\lambda_5+\alpha_6\lambda_6+\alpha_7\lambda_7  \right)}.
\end{equation}

Notice that $\lambda_3,\lambda_8,$ (and $\lambda_0$ if it is used) are diagonal matrices and represent the phases.  $\lambda_1,\lambda_2$ are off-diagonal matrices and represent the transitions between the ground and first excited states.  $\lambda_6,\lambda_7$ are off-diagonal matrices and would represent the transitions between the first and second excited states.  $\lambda_4,\lambda_5$ are off-diagonal matrices and would represent the transitions between the ground and second excited states.
\label{sec:OtherPs}

Note that in Eq.~(\ref{eq:cd}) $\lambda_1$, and $\lambda_2$ seem to be different from $\lambda_4,\lambda_5,\lambda_6,\lambda_7$, even though the three qutrit states are fundamentally equivalent and can be permuted. However, there is a difference arising from the choice of $\lambda_3$ and $\lambda_8$ as diagonal matrices.  One needs a complete set of traceless, Hermitian matrices and other choices are also acceptable.  One could instead choose any one of the three following matrices as the first diagonal matrix:
\begin{equation}
\left(\begin{array}{ccc}  1 & 0 & 0 \\ 0 & -1 & 0 \\ 0 & 0 & 0 \end{array}\right),
\left(\begin{array}{ccc}  1 & 0 & 0 \\ 0 & 0 & 0 \\ 0 & 0 & -1 \end{array}\right),
\left(\begin{array}{ccc}  0 & 0 & 0 \\ 0 & 1 & 0 \\ 0 & 0 & -1 \end{array}\right).
\end{equation}
The choice would then determine the form of the other independent diagonal matrix. Specifically, the corresponding matrix would be 
\begin{equation}
 \frac{1}{\sqrt{3}}\left(\begin{array}{ccc}  1 & 0 & 0 \\ 0 & 1 & 0 \\ 0 & 0 & -2 \end{array}\right),
\frac{1}{\sqrt{3}}\left(\begin{array}{ccc}  1 & 0 & 0 \\ 0 & -2 & 0 \\ 0 & 0 & 1 \end{array}\right),
\frac{1}{\sqrt{3}}\left(\begin{array}{ccc}  -2 & 0 & 0 \\ 0 & 1 & 0 \\ 0 & 0 & 1 \end{array}\right),
\end{equation}
respectively. The choice of diagonal matrices in Eq.~(\ref{eq:cd}) leads to three different but equivalent parametrizations.

Here, the Cartan decomposition  \cite{helgason1979differential, hermann1966lie} is used to decompose any unitary matrix.  
Let $\la{G}$ be the Lie algebra of a Lie group $G$.  Let $\la{K}$ and $\la{P}$ be subsets of $\la{G}$.  The set $\la{G}$ is a complete set of traceless, Hermitian $3\times 3$ matrices. They are complete in the sense that any traceless, Hermitian $3\times 3$ matrix can be expressed as a real, linear combination of the basis set.  Recursive Cartan decomposition has also been done in \cite{jiang2023optimalsynthesisgeneralmultiqutrit}, which reduced the number of gates compared to the earlier work \cite{PhysRevA.92.062317, PhysRevA.87.012325}.

Let $p_i$ be any element of $\la{P}$ and $k_j$ be any element of $\la{K}$.  Now, suppose that the following commutation relations hold:
\begin{equation}\label{eq:cdc}
[k_i,k_j] \in \la{K}, \;\; [p_i,p_j] \in \la{K}, \;\;\; [p_i,k_j]\in \la{P}. \end{equation}
That is, if we commute any two elements from the set $\la{K}$, we get another element of $\la{K}$. If we commute any two elements of $\la{P}$ we also get an element of $\la{K}$. If we take the commutator of one element of $\la{P}$ and one element of $\la{K}$, then we get an element of $\la{P}$.

Then an element of the group $G$ can be written as 
$$
G= K\cdot P,
$$
where $K$ is the exponential of $\la{K}$ and $P$ is the exponential of $\la{P}$.

Let $\la{SU(3)}$ be the Lie algebra of $SU(3)$.  Let $\la{K}$ and $\la{P}$ be defined as follows:
$$
\la{P} =\{\lambda_4,\lambda_5,\lambda_6, \lambda_7\}, \mbox{ and } \la{K} = \{\lambda_1,\lambda_2,\lambda_3,\lambda_8\},
$$
where the $\{\lambda_i\}$ are the Gell-Mann matrices.  
Since the commutation relations above are satisfied, we have 
$$
G = K\cdot P,
$$ 
where $G$ is $SU(3)$ and $K$ and $P$ are defined above.  

$K$ can further be decomposed into an $SU(2)$ part and $\lambda_8$ part because $\lambda_8$ commutes with $\lambda_1, \lambda_2$ and $\lambda_3$.  Then an Euler angle decomposition to $SU(2)$ can be used (which is actually also a Cartan decomposition with $\lambda_3 = \la{K}$).  The group element can then be written as 
$$
G=e^{-i\lambda_8\alpha_8}e^{-i\lambda_3\alpha_1}e^{-i\lambda_2\alpha_2}e^{-i\lambda_3\alpha_3}P
$$
The objective is to write the diagonal elements to the left and the off-diagonal elements to the right.  To do this, insert the identity in the form 
$$
e^{-i\lambda_3\alpha_3}e^{+i\lambda_3\alpha_3} = \id,
$$
so 
$$
G=e^{-i\lambda_8\alpha_8} e^{-i\lambda_3\alpha_1} e^{-i\lambda_3\alpha_3} e^{+i\lambda_3\alpha_3} e^{-i\lambda_2\alpha_2} e^{-i\lambda_3\alpha_3}P.
$$
Now, rewrite the following factor
$$
e^{+i\lambda_3\alpha_3} e^{-i\lambda_2\alpha_2}e^{-i\lambda_3\alpha_3}.
$$
Let $V = e^{+i\lambda_3\alpha_3}$, and note that 
$$
Ve^{-i\lambda_2\alpha_2} V^\dagger = e^{-iV\lambda_2 V^\dagger\alpha_2}.
$$
Now
$$
V\lambda_2 V^\dagger = \lambda_2\cos(2\alpha_3) +\lambda_1\sin(2\alpha_3).
$$
Now, define the following new parameters 
$$
\beta_1 \equiv \alpha_2\sin(2\alpha_3) \;\; \beta_2 \equiv \alpha_2\cos(2\alpha_3) \;\; \beta_3 \equiv \alpha_1+\alpha_3
$$
This can all be put back into $G$ to arrive at the following expression for an element of $SU(3)$.
After this, we have the following decomposition
$$
G= U_d\cdot U_{o1} \cdot U_{o2},
$$
where $U_d$ is diagonal, and $U_{o1}$, $U_{o2}$ are exponentials of only off-diagonal terms.  

Explicitly, the final form is 
\begin{eqnarray}\label{CD1}
G &=& \exp\left[-i\left(\beta_3\lambda_3+\alpha_8\lambda_8  \right)\right]\times \exp\left[-i\left(\beta_1\lambda_1+\beta_2\lambda_2  \right)\right] \nonumber \\
&&\times\exp\left[-i\left(\alpha_4\lambda_4+\alpha_5\lambda_5+\alpha_6\lambda_6+\alpha_7\lambda_7  \right)\right],
\end{eqnarray}
where all $\alpha_i$ and $\beta_j$ are real parameters.

The form of this decomposition, as is the commutation relations and anti-commutation relations are determined by the choice of basis.  The basis could be chosen as 
\begin{equation}
\la{G^\prime} \equiv \{
\lambda_1,\lambda_2,\lambda_4,\lambda_5,\lambda_6,\lambda_7,\lambda_9,\lambda_{11}\}
\end{equation}
or 
\begin{equation}
\la{G^{\prime\prime}}\equiv \{
\lambda_1,\lambda_2,\lambda_4,\lambda_5,\lambda_6,\lambda_7,\lambda_{10},\lambda_{12}\},
\end{equation}
where the matrices $\{\lambda_9,\lambda_{10},\lambda_{11},\lambda_{12}\}$ can be substitutions for the diagonal basis elements $\lambda_3$ and $\lambda_8$, which are given in Appendix \ref{sec:AppendixA}.   
Then there are different choices for the subsets $\la{K}$, and $\la{P}$.  Let us choose a set 
$$
\la{P^\prime} =\{\lambda_1,\lambda_2,\lambda_4, \lambda_5\}, \mbox{ and } \la{K^\prime} = \{\lambda_6,\lambda_7,\lambda_{10},\lambda_{12}\},
$$
and 
$$
\la{P^{\prime\prime}} =\{\lambda_1,\lambda_2,\lambda_6, \lambda_7\}, \mbox{ and } \la{K^{\prime\prime}} = \{\lambda_4,\lambda_5,\lambda_9,\lambda_{11}\}.  
$$
Following the same procedure as above, with $\la{K}$, and $\la{P}$ replaced by their  counterparts, the following two expressions can be obtained:
\begin{eqnarray}\label{eq:cd_2}
G &=& e^{-i\left(\beta_{10}\lambda_{10}+\alpha_{12}\lambda_{12}  \right)}e^{-i\left(\beta_6\lambda_6+\beta_7\lambda_7  \right)} \nonumber \\
&&\times e^{-i\left(\alpha_4\lambda_4+\alpha_5\lambda_5+\alpha_1\lambda_1+\alpha_2\lambda_2  \right)},
\end{eqnarray}
and 
\begin{eqnarray}\label{eq:cd_3}
G &=& e^{-i\left(\beta_9\lambda_9+\alpha_{11}\lambda_{11}  \right)}e^{-i\left(\beta_4\lambda_4+\beta_5\lambda_5  \right)} \nonumber \\
&&\times e^{-i\left(\alpha_1\lambda_1+\alpha_2\lambda_2+\alpha_6\lambda_6+\alpha_7\lambda_7  \right)}
\end{eqnarray}
respectively.

Since, as mentioned above, two-photon transitions can be more difficult to implement than single-photon, dipole, transitions, it can be desirable to eliminate the elements $\lambda_4$ and $\lambda_5$ that correspond to two-photon transitions \cite{ByrdSU3, PhysRevA.74.032334}. The two new decompositions in Eq.~(\ref{eq:cd_2}) and Eq.~(\ref{eq:cd_3}) are physically different from Eq.~(\ref{eq:cd}), even though they are mathematically equivalent (in the sense that they both parameterize the group $S(3)$). This new decomposition therefore presents another approach to qutrit gate decompositions based on experimental considerations.

\subsection{Givens Rotations}

 It is worth mentioning that given a three-level system, implementing an operation initially on the first two states ($\ket{0}$ and $\ket{1}$), followed by an operation on the second two states ($\ket{1}$ and $\ket{2}$) and another operation on the first two states ($\ket{0}$ and $\ket{1}$), allows us to do any operation on qutrits. These operations are elementary rotations in a 2-dimensional subspace of a larger space and assist in representing or approximating a unitary matrix \cite{Givens1958ComputationOP, Nielsen, Frerix2019ApproximatingOM, PhysRevLett.92.177902}. The middle factor of Eq.~(\ref{eq:cd_3}) can be eliminated, removing two-photon transitions, using the following relations:
\begin{eqnarray}
e^{i\lambda_i\theta}\lambda_je^{-i\lambda_i\theta} &=& \lambda_j + i\theta[\lambda_i,\lambda_j] +\frac{1}{2!}\left[\lambda_i,\left[\lambda_i,\lambda_j\right]\right] \nonumber \\
&& + \frac{-i}{3!}\left[\lambda_i,\left[\lambda_i,\left[\lambda_i,\lambda_j\right]\right]\right] + \cdots
\end{eqnarray}
and
\begin{equation}
    Ue^{-i\lambda_i\theta}U^\dagger = e^{-iU\lambda_iU^\dagger\theta}.
\end{equation}
The first of these can be used to show that 
\begin{equation}\label{eq:lam5}
e^{i\lambda_2\theta}\lambda_5e^{-i\lambda_2\theta} = \lambda_5\cos\theta +\lambda_7\sin\theta
\end{equation}

$$
e^{i\lambda_2\theta} = \left(\begin{array}{ccc}
0&0&0\\
0&0&0\\
0&0&1 
\end{array}\right)
+\lambda_2(i\sin\theta) +
\left(\begin{array}{ccc}
1&0&0\\
0&1&0\\
0&0&0 
\end{array}\right)\cos\theta. 
$$
The same thing can be applied to $\lambda_4$
\begin{equation}\label{eq:lam4}
e^{i\lambda_2\theta}\lambda_4e^{-i\lambda_2\theta} = \lambda_4\cos\theta +\lambda_6\sin\theta.
\end{equation}
Using Eq.~(\ref{eq:lam5}), and Eq.~(\ref{eq:lam4}) , Eq.~(\ref{eq:cd_3}) can be written as:
\begin{eqnarray}\label{eq:wthe}
G &=& e^{-i\left(\beta_9\lambda_9+\alpha_{11}\lambda_{11}  \right)}\nonumber\\
&&\times e^{-i\left(\beta_4(\lambda_4\cos\theta +\lambda_6\sin\theta)+\beta_5(\lambda_5\cos\theta +\lambda_7\sin\theta)  \right)} \nonumber \\
&&\times e^{-i\left(\alpha_1\lambda_1+\alpha_2\lambda_2+\alpha_6\lambda_6+\alpha_7\lambda_7  \right)}.
\end{eqnarray}  
Assuming that $\theta=\frac{\pi}{2}$ we can write Eq.~(\ref{eq:wthe}) as
\begin{eqnarray}\label{eq:Gfin}
G &=& e^{-i\left(\beta_9\lambda_9+\alpha_{11}\lambda_{11}  \right)}e^{-i\left(\beta_4\lambda_6+\beta_5\lambda_7  \right)} \nonumber \\
&&\times e^{-i\left(\alpha_1\lambda_1+\alpha_2\lambda_2+\alpha_6\lambda_6+\alpha_7\lambda_7  \right)}
\end{eqnarray}
where $\lambda_4$ and $\lambda_5$ were replaced with $\lambda_6$ and $\lambda_7$ respectively.

The group $SU(3)$ was analyzed in detail in \cite{ByrdSU3,ByrdSU3E}.  In that work, the group was decomposed into an "Euler angle" decomposition, meaning all $U(1)$ subgroups.  The decomposition used $K$ and $P$ as in Eq.~\ref{eq:cd}.  The result was 
\begin{equation}
    e^{i\lambda_3\alpha}e^{i\lambda_2\beta}e^{i\lambda_3\gamma}e^{i\lambda_5\theta}e^{i\lambda_3 a}e^{i\lambda_2 b}e^{i\lambda_3 c}e^{i\lambda_8\phi}.
\end{equation}
Note that there are three off-diagonal matrix transformations which correspond to 3 Givens rotations.  (Note that Murnaghan also gave a decomposition similar to Givens, \cite{Murnaghan}.)  If $K$ and $P$ are chosen to be 
$$
\la{P^\prime} =\{\lambda_1,\lambda_2,\lambda_4, \lambda_5\}, \mbox{ and } \la{K^\prime} = \{\lambda_6,\lambda_7,\lambda_{10},\lambda_{12}\},
$$
then the group element can be chosen to be
\begin{equation}
    e^{i\lambda_{10}\alpha}e^{i\lambda_6\beta}e^{i\lambda_{10}\gamma}e^{i\lambda_2\theta}e^{i\lambda_{10} a}e^{i\lambda_6 b}e^{i\lambda_{10} c}e^{i\lambda_{12}\phi}.
\end{equation}
Note that, excluding the phases, this consists of a coupling between the ground and first excited state, followed by coupling the first excited state and the second excited state, followed by a coupling between the ground and first excited state.  In other words, it shows the ability to perform the transformations stated in the first paragraph of this section.  

\subsection{Walsh-Hadamard}

Using the set of parameters in Eq.~(\ref{eq:Gfin}), the Walsh-hadamrd matrix is obtained. Note that Walsh-Hadamrd is a unitary matrix but G in Eq.~(\ref{eq:Gfin}) is a special unitary matrix which by adding $\lambda_0$ as the identity matrix will be converted to a unitary matrix.

The Hamiltonians related to  Eq.~(\ref{eq:Gfin}) when rewriting it as 
$G = e^{-iH_1\theta_1}\times e^{-iH_2\theta_2}\times e^{-iH_3\theta_3}$
are
\begin{equation*}
    H_1 = \left(\begin{array}{ccc}
        0.9631 & 0 & 0 \\
        0 & -0.6091 & 0 \\
        0 & 0 & 0.8471
    \end{array}\right)
\end{equation*}
with $\theta_1=6.5239$,
\begin{equation*}
 H_2= \left(\begin{array}{ccc}
        0 & 0 & 0 \\
        0 & 0 & 1 \\
        0 & 1 & 0
    \end{array}\right)
\end{equation*}
with $\theta_2=5.9977$ and
\begin{equation*}
 H_3 = \left(\begin{array}{ccc}
        0 & -0.5907i & 0 \\
        0.5907i& 0& -0.8069i \\
        0 & 0.8069i & 0
    \end{array}\right)
\end{equation*}
with $\theta_3=4.4994$. 
We can also use another type of Cartan decomposition, which is not just physically but also mathematically different from Eq.~(\ref{eq:cd}) by defining $\la{K}$ and $\la{P}$ as follows:
$$
\la{P} =\{\lambda_1, \lambda_3, \lambda_4,\lambda_6,\lambda_8\}, \mbox{ and } \la{K} = \{\lambda_2,\lambda_5,\lambda_7\}.
$$
The same commutation relations hold for these 2 sets, meaning:
$$
[k_i,k_j] \in \la{K}, \;\; [p_i,p_j] \in \la{K}, \;\;\; [p_i,k_j]\in \la{P}, 
$$
shown in Table \ref{tbl:tbl1} where $p_i$ are any element of $\la{P}$ and $k_j$ are any element of $\la{K}$.


\bibliography{ecc}

\end{document}